\documentclass[12pt]{article}
\usepackage{graphicx}
\begin{document}
\baselineskip=5.5mm
\newcommand{\be} {\begin{equation}}
\newcommand{\ee} {\end{equation}}
\newcommand{\Be} {\begin{eqnarray}}
\newcommand{\Ee} {\end{eqnarray}}
\def\a{\alpha}
\def\b{\beta}
\def\g{\gamma}
\def\G{\Gamma}
\def\d{\delta}
\def\D{\Delta}
\def\e{\epsilon}
\def\k{\kappa}
\def\l{\lambda}
\def\s{\sigma}
\def\t{\tau}
\def\om{\omega}
\def\Om{\Omega}
\def\lg{\langle}
\def\rg{\rangle}

\noindent
\begin{center}
{\Large
{\bf
Dynamic heterogeneities in the out-of-equilibrium dynamics of simple
spherical spin models\\
}}
\vspace{1cm}
\noindent
{\bf Gregor Diezemann} \\
{\it
Institut f\"ur physikalische Chemie, Universit\"at Mainz,
Welderweg 11, 55099 Mainz, FRG
\\}
revised version\\
\end{center}
\vspace{2cm}
\noindent
{\it
The response of spherical two-spin interaction models, the spherical
ferromagnet (s-FM) and the spherical Sherrington-Kirkpatrick (s-SK) model, is
calculated for the protocol of the so-called nonresonant hole burning
experiment (NHB) for temperatures below the respective critical temperatures.
It is shown that it is possible to select dynamic features in the
out-of-equilibrium dynamics of both models, one of the hallmarks of
dynamic heterogeneities.
The behavior of the s-SK model and the s-FM model in three dimensions is very
similar, showing dynamic heterogeneities in the long time behavior, i.e. in
the aging regime.
The appearence of dynamic heterogeneities in the s-SK model explicitly
demonstrates that these are not necessarily related to {\it spatial}
heterogeneities.
For the s-FM it is shown that the nature of the dynamic heterogeneities changes
as a function of dimensionality. With incresing dimension the frequency
selectivity of the NHB diminishes and the dynamics in the
mean-field limit of the s-FM model becomes homogeneous.
}

\vspace{1cm}
\vspace{1.5cm}
\section*{I. Introduction}
Non-exponential relaxation behavior is found to be rather common when dealing
with disordered materials like glasses, spin-glasses, disordered crystals
or proteins\cite{a.rel}.
In the last decade particular attention has been payed to the question
to which extent the relaxation is to be viewed as dynamic
heterogeneous\cite{het.dyn}.
Different experimental techniques have been invented in order to investigate
the detailed nature of the relaxation particularly of amorphous
systems\cite{SRS91, CE95, ViRI00, DvB01}.
These techniques allow to specifically select a slow sub-ensemble and
afterwards monitor its relaxation.
Various applications of these techniques have shown that the primary response
in amorphous polymers and supercooled liquids is to be viewed as
heterogeneous in the sense that it is possible to select slow sub-ensembles
relaxing at smaller rates than the average.
Throughout this paper I use the definition given in ref.\cite{het.def}
according to which a system will be called dynamic heterogeneous if it is
possible to select dynamically distinguishable (slow and/or fast)
contributions to a relaxation.

In addition it has been found that after a certain re-equilibration time the
relaxation properties of the selected sub-ensemble return to those of the
bulk\cite{andreas.4d, roland.4d}.
Therefore, these experiments indicate that the response can be described
as a superposition of exponentially decaying entities with different
relaxation rates. The various relaxation rates, however, are not static
quantities but apparently fluctuate in time. Different interpretations
have been provided for this behavior\cite{heuer97, dieze97}.
The NMR-techniques\cite{andreas.4d, roland.4d} have the advantage of a simple
interpretation in terms of equilibrium 4-time correlation functions but are
restricted to a rather narrow temperature regime and to certain materials.
The optical deep bleach technique\cite{CE95} has the advantage that it can be
applied in a wider temperature range with the shortcoming that up to now it has
not been interpreted in terms of equilibrium correlation functions.
At this point it is important to note that these experimental methods monitor
molecular reorientations and therefore are {\it not} able to address the
question of {\it spatial aspects} of the dynamic heterogeneities.
The only exception is provided by a new variant of the 4d-NMR
technique\cite{4dcp}, which allows to extract a length scale via monitoring
spin-diffusion. Originally, this technique has been applied to a polymeric
liquid and later on the length scale of the dynamic heterogeneities has
been extracted for low-molecular glass-forming systems\cite{4dcp.gly,4dcp.vgl}.
The corresponding length scales have been found to be on the order of
$1\cdots 4$nm.

Another experimental technique allowing to monitor dynamic heterogeneities
is provided by the nonresonant spectral hole burning (NHB)
experiment\cite{SBLC96}.
This method is based on a pump-wait-probe field sequence with a large pump
field amplitude beyond the linear response regime.
Though originally applied to supercooled liquids\cite{SCDB97}, NHB in the
meantime has been used to investigate the relaxation of several materials,
including disordered crystals (relaxor ferroelectrics)\cite{KSB98},
ion-conducting glasses\cite{RB99} and spin glasses\cite{ralph99}.
Also the application of NHB to a solvable glass-model has been
presented\cite{leticia01}.
The interpretation of the obtained results has mainly been guided by the fact
that via the application of a large amplitude ac field of frequency $\Om$
the sample absorbs energy of an amount proportional to the imaginary part
of the susceptibility evaluated at the pump frequency $\Om$\cite{kubo}.
In case of a homogeneously broadened response one does not expect that it
is possible to modify the response in a frequency selective way. By contrast
such a goal could be achieved if the response is given by a superposition of
differently fast relaxing entities (heterogeneous scenario). This is because
in this case energy absorbtion should be largest for those sub-ensembles with
a relaxation time on the scale of the inverse pump frequency.
This intuitive picture is confirmed in the framework of a response-theory for
NHB for the particular case of stochastic dipole reorientations, which I have
developed recently\cite{dieze01}.

In the quoted experiments it always has been found that a frequency selective
modification of the response indeed is possible. Regarding the
re-equilibration of this modification, however, the results differ not only
with respect to the frequency dependence but also regarding the time scale
of the recovery. For the latter a time scale longer than the inverse
burn frequency has been observed in the case of the relaxor
ferroelectrics\cite{KSB98}.

As long as one is concerned with supercooled liquids, one can safely consider
the system to be in (metastable) equilibrium prior to the NHB field sequence.
This, however, is not necessarily true for the relaxor materials or the spin
glasses.
In particular, it was argued in ref.\cite{DB01} that the results obtained for
the solvable p-spin-glass model\cite{leticia01} are mainly to be interpreted
in terms of out-of-equilibrium effects.
In equilibrium, the $p\!=\!3$-model studied shows an exponential relaxation
at long times. Therefore, according to what was said above one does not expect
to be able to select a sub-ensemble in a frequency dependent way in the
equilibrated version of the model.
It has to be mentioned here that dynamic heterogeneities as monitored by
NHB have also been observed in Monte Carlo simulations on an equilibrated
Sherrington-Kirkpatrick mean-field spin-glass model\cite{let.reply}.
These calculations explicitly demonstrate that from the observation of
dynamic heterogeneities one cannot conclude on the existence of spatial
heterogeneities.
Also, this finding appears to be independent of whether the system was in
thermal equilibrium before the application of the pump-field.

In this paper I consider the application of the NHB field sequence to the
spherical Sherrington-Kirkpatrick (s-SK) model, i.e. the p-spin model with
$p=2$ and the spherical ferromagnet (s-FM) in arbitrary dimension.
The Langevin dynamics for these models has been solved analytically\cite{CD95}.
Additionally, the s-FM model is equivalent to the O(N)-model in the limit of
large N and therefore is a typical model for domain coarsening
processes\cite{bray94}.

I will solely consider a thermal history protocol in which the system is
quenched to a temperature below the critical temperature $T_c$ from infinite
temperature prior to the experiment. The behavior at and above $T_c$ will be
investigated in a forthcoming publication.
Therefore, all effects observed are intimately related to the
out-of-equilibrium dynamics of the model.
This is because the system never reaches equilibrium in this temperature
regime.
The outline of the paper is as follows. In the next Section I will briefly
recall the dynamic features of the models and calculate the response to
the NHB field sequence in second order regarding the pump-field amplitude
and linearly in the small step-field.
In Section III the results of the calculations are presented and discussed.
The paper closes with some conclusions in Section IV.
\section*{II. NHB in spherical two-spin interaction models}
The spherical models under consideration are defined by the Hamiltonian
\be\label{Ham.def}
{\bf H}=-\frac{1}{2}\sum_{i\neq k}J_{ik} s_i s_k-\sum_i h_i s_i
\ee
where in case of the s-SK model the $J_{ik}$ are chosen at random from a
Gaussian probability distribution with zero mean and variance $\s=1/N$
and are restricted to a ferromagnetic coupling $J$ (to be set to unity in the
following) for the s-FM model on a simple hypercubic lattice in d dimensions.
In addition the spin-variables are subject to the spherical constraint
$\sum_i s_i^2=N$.
The Langevin equations governing the dynamics of the model read as
\be\label{Langevin}
{\dot s}_i(t)=\sum_k J_{ik}s_k(t)+h_i-z(t)s_i(t)+\xi_i(t)
\ee
where $z(t)$ is the Lagrange multiplier enforcing the spherical constraint
and $\xi_i(t)$ is a $\d$-correlated Gaussian white noise.
These equations have been solved analytically by Cugliandolo and Dean
(CD)\cite{CD95} in two papers.
The dynamical properties of the s-FM model are discussed in ref.\cite{GL00}.
Further information regarding the correspondence between the two models can be
found in ref.\cite{ZKH00}. Here, I briefly summarize the results relevant in
the present context.

Of particular importance are the violations of the fluctuation-dissipation
theorem (FDT), which relates the response function to the time-derivative of
the two-time correlation function,
\be\label{FDT}
R(\t)=-{1\over T}{dC(\t)\over d\t}
\ee
In particular, it has proven extremely useful in out-of -equilibrium
situations to define the so-called fluctuation-dissipation ratio $X(t,t_w)$
via\cite{FDT_rev98},
\be\label{X_FDT}
R(t,t_w)={ X(t,t_w)\over T}{\partial C(t,t_w)\over\partial t_w}
\ee
the limiting long-time behavior, $X_\infty$, of which is known to vanish for
domain coarsening models\cite{X_infty}. Here, $t_w$ denotes the time that has
elapsed after a quench to the working temperature prior to the measurement.
Also for the models considered in the present paper one has
$X_\infty=0$\cite{GL00}.
Thus, concerning this measure of typical distances
from equilibrium, the domain coarsening models in finite dimension do not show
any differences to the mean-field s-SK model.

The response of the system, $R_h(t,t')=\sum_i^N\lg s_i(t)\xi_i(t')\rg/(2NT)$,
in the presesence of a field $h(t)$ can be obtained in the same way as
calculated by CD for the zero field case.
As shown by Berthier et al.\cite{BCI01} this yields:
\be\label{R.t.tw}
R_h(t,t')=\theta(t-t')\frac{W_h(t')}{W_h(t)}g\!\left(\frac{t-t'}{2}\right)
\ee
where the function $g(t)$ is defined by
\Be\label{g.t.def}
&&g(t)=\left[\exp{(-4t)}I_0(4t)\right]^{d}\quad \mbox{s-FM}\nonumber\\
&&g(t)=\exp{(-4t)}\frac{I_1(4t)}{2t}\quad \mbox{s-SK}
\Ee
with $I_n(x)$ denoting the generalized Bessel function and $W_h(t)$ is the
solution of
\Be\label{Wh.def}
W_h(t)^2=
&&\hspace{-0.6cm}
g(t)+2T\int_0^t\!d\t W_h(\t)^2g(t-\t)
\nonumber\\
&&\hspace{-0.4cm}
+\int_0^t\!dt_1\int_0^{t}\!dt_2h(t_1)h(t_2)W_h(t_1)W_h(t_2)
			g\!\left(t-\frac{t_1+t_2}{2}\right)
\Ee
which has its origin in the normalization of the equal-time correlation,
$C(t,t)=1$, i.e. the spherical constraint.

\subsection*{Zero field response}

Before I turn to the calculation of the response following the NHB pulse
sequence, it is appropriate to summarize the known results for the reponse
and the correlation in zero field, for a more detailed discussion see
refs.\cite{CD95,GL00,ZKH00}.
In zero field, eq.(\ref{Wh.def}) simplifies to the following Volterra equation:
\be\label{W.t}
W(t)^2=g(t)+2T\int_0^t\!d\t W(\t)^2g(t-\t)
\ee
For the s-SK model, this equation has been solved by CD for random initial
conditions (i.e. a quench from $T=\infty$ at $t=0$).
No simple analytical solution exists for the s-FM model.
However, for $t\gg 1$ it can be shown\cite{CD95, GL00, ZKH00} that for
$T<T_c$
\be\label{W.as}
W(t)^2=\frac{1}{(1-T/T_c)^2}g_{as}(t)
\ee
Here, the asymptotic behavior of the $g(t)$ are given by
\Be\label{g.as.def}
&&g_{as}(t)=(8\pi t)^{-d/2}\hspace{1.3cm}\mbox{s-FM}\nonumber\\
&&g_{as}(t)=(32\pi)^{-1/2}t^{-3/2}\quad\mbox{s-SK}
\Ee
The critical temperatures are given by $T_c=1$ for the s-SK model and
$T_c$ depends on the spatial dimension in case of the s-FM model,
with $T_c(d=3)\simeq 3.9568$\cite{BK52}. For the other dimensions used in the
present paper I find $T_c(d=5)\simeq 8.6482$, $T_c(d=7)\simeq 12.7982$ and
$T_c(d=9)\simeq 16.8579$.

If $t_w$ denotes the time elapsed after a quench from $T=\infty$ to $T<T_c$,
the following behavior is found for $R(t+t_w,t_w)$ from
eqns.(\ref{R.t.tw},\ref{W.as},\ref{g.as.def}) in the limit of long $t_w$:
\Be\label{R.as}
&&R(\t+t_w,t_w)
   =\l^{-\e/4}g(\t/2)
   \quad\mbox{with}\quad
   \l=\frac{t_w}{t_w+\t} \quad ;  \quad t_w \gg 1
      \nonumber\\
   &&\hspace{1cm} \e=d\quad\mbox{(s-FM)}\quad ; \quad \e=3\quad\mbox{(s-SK)}
\Ee
From this expression it is evident already that the dynamic properties of the
s-SK model are very similar to those of the s-FM in $d=3$.
Therefore, the following expressions are given for the s-FM model. The only
differences between the s-FM model in $d=3$ and the s-SK model stem from the
different prefactors in eq.(\ref{g.as.def}).
In particular, it is evident from eq.(\ref{R.as}) that the temperature is an
irrelevant variable in the whole low-temperature phase.

Two time-sectors are to be distinguished:
\begin{itemize}
\item
For short times $\t$ such that $\t\ll t_w$ - the so-called stationary regime -
one has $\l\simeq 1$ and accordingly:
\be\label{R.stat}
R(\t+t_w,t_w) = R(\t) = g(\t/2)\quad \t\ll t_w
\ee
In particular, in this stationary regime the FDT, eq.(\ref{FDT}), holds.
In the long time limit, $1\ll \t$, eq.(\ref{R.stat}) shows that $R(\t)$
decays according to $R(\t)\sim \t^{-d/2}$.
\item
In the so-called aging regime $1\ll \t\sim t_w$, one finds from eq.(\ref{R.as}):
\be\label{R.aging}
R(\t+t_w,t_w)=\l^{-d/4}(4\pi \t)^{-d/2} \quad 1\ll \t\sim t_w
\ee
In this regime the FDT is strongly violated.
If in addition $\t\gg t_w$, the response behaves as $R(\t,t_w)\sim \t^{-d/4}$.
\end{itemize}

The overall behavior is shown for various values of the waiting time in Fig.1.
From this plot it is seen that for all dimensions shown, $d=3,5,7$, there is a
crossover from the $\t^{-d/2}$ to the $\t^{-d/4}$ behavior. Also the explicit
dependence on the waiting time in the aging regime is evident.
It should be pointed out that the crossover from the stationary regime to the
aging (domain growth) regime takes place around $\t\sim t_w$ independent of
spatial dimension.
Remember, that only in the stationary regime the FDT holds.
In particular, the behavior of the response (and also the two-time correlation
function) does not change qualitatively around $d\!=\!4$, above which the
model behaves mean-field like concerning the statics and the exponents.
(The same holds for the dynamic fluctuations, as will be shown elsewhere.)

The above discussion shows that the relaxation is extremely non-exponential.
Therefore, the question as to what extent the response can be viewed as
dynamic heterogeneous naturally arises.
\subsection*{\bf Nonresonant hole burning}
In the following, the response will be calculated for the NHB-field-sequence,
cf. Fig.2.
At $t=0$ the system is quenched from $T=\infty$ to the working temperature $T$.
One (or more) cycles of the pump-field $h_p(t)=h_p\sin{[\Om(t-t_q)]}$ are
applied after a time $t_q$ has elapsed.
Following a waiting time $t_w$ the response $R^*(\hat{t}+\t,\hat{t}+\t')$ is
measured, where I defined
\[\hat{t}=t_q+t_p+t_w\quad\mbox{and}\quad t_p=2N\pi/\Om\]
for brevity. Here, $N$ denotes the number of cycles of the sinusoidal
pump-field.
In the following calculations of $R^*$ the $W_h(t)$ are needed in second
order with respect to the pump-field amplitude $h_p$.
From eq.(\ref{Wh.def}) it is evident that a perturbation expansion follows
from
\be\label{W.hp}
W_{h_p}(t)^2=W(t)^2+h_p^2 \D(t)^2+{\cal O}(h_p^4)
\ee
Inserting this expression into eq.(\ref{Wh.def}) yields eq.(\ref{W.t}) for
the zero'th order term and, assuming a time-dependent field of the form
$h(t)=h_p\sin{[\Om(t-t_q)]}\theta(t-t_q)$ according to Fig.2:
\Be\label{del.W}
&&\hspace{-1.0cm}\D(t_q+\t)^2=
2T\int_0^\t\!ds \D(t_q+s)^2g(\t-s) + \D_0(t_q+\t)
\nonumber\\
&&\hspace{-1.0cm}
\D_0(t_q+\t)=\int_0^{t_m}\!dt_1\int_0^{t_m}\!dt_2\sin{(\Om t_1)}\sin{(\Om t_2)}
			W(t_q+t_1)W(t_q+t_2)
			g\!\left(\t-\frac{t_1+t_2}{2}\right)
\Ee
where $t_m=$Min$(\t,t_p)$.
Using eq.(\ref{W.hp}) the response function in ${\cal O}(h_p^2)$ then is found
to be given by ($\t>\t'$):
\Be\label{R.DR.allg}
&&\hspace{-0.8cm}R^*(\hat{t}+\t,\hat{t}+\t') = R(\hat{t}+\t,\hat{t}+\t')
+\D R(\hat{t}+\t,\hat{t}+\t')
\nonumber\\
&&\hspace{-0.8cm}R(\hat{t}+\t,\hat{t}+\t')
   =\frac{W(\hat{t}+\t')}{W(\hat{t}+\t)}g\!\left(\frac{\t-\t'}{2}\right)\\
&&\hspace{-0.8cm}\D R(\hat{t}+\t,\hat{t}+\t')=
-\frac{h_p^2}{2}
\left[\frac{\D(\hat{t}+\t)^2}{W(\hat{t}+\t)^2}-
 \frac{\D(\hat{t}+\t')^2}{W(\hat{t}+\t')^2}\right]
R(\hat{t}+\t,\hat{t}+\t')\nonumber
\Ee
In the NHB-protocol of Fig.2, however, the response to a small step field is
recorded, i.e. the integrated response (the thermoremanent magnetization):
\be\label{Chi.def}
\chi^*(\hat{t},\t)=\chi(\hat{t},\t)+\D\chi(\hat{t},\t)
=\int_0^\t\!ds R^*(\hat{t}+\t,\hat{t}+s)
\ee
according to eq.(\ref{R.DR.allg}) with the zero-field integrated response
$\chi(\hat{t},\t)=\int_0^\t\!ds R(\hat{t}+\t,\hat{t}+s)$.
Eqns.(\ref{del.W})$\cdots$(\ref{Chi.def}) allow the calculation of the results
of a NHB experiment at any desired temperature.
\section*{III. Results and discussion}
Many of the general features of the modification $\D\chi(\hat{t},\t)$ can
already be seen for $T=0$, which is the simplest case.
Afterwards finite temperatures will be discussed as well as the dependence
of the observed features on spatial dimension. Finally, a direct comparison
between the three-dimensional s-FM model and the s-SK model will by carried
out.

Throughout the remainder of the paper the dependence on $\hat{t}$ will be
skipped whenever no confusion can occur, i.e. the shorthand notation
$\chi(\t)\equiv\chi(\hat{t},\t)$ and $\D\chi(\t)\equiv\D\chi(\hat{t},\t)$ will
be used. Times and frequencies will be given in dimensionless units.
\subsection*{A. Spherical ferromagnet}
In this subsection the details of the results for calculations of the
response following the NHB field sequence are discussed.
The discussion is kept general with regard to spatial dimension but the
actual calculations are carried out for $d\!=\!3$, cf. Figs.3$\cdots$6.
The dependence of the results on spatial dimension will be presented in the
next subsection.
\subsubsection*{T=0:}
For $T\!=\!0$, eq.(\ref{del.W}) can be solved trivially and the modified
response is easily calculated. From eq.(\ref{W.t}) one explicitly has
\be\label{W.del.T0}
W(t)^2=g(t)\quad\mbox{and}\quad \D(t)^2=\D_0 (t)\quad ; \quad T=0
\ee
The corresponding expressions for $\chi(\t)$ and $\D\chi(\t)$ are easily
obtained from eqns.(\ref{R.DR.allg},\ref{Chi.def}).

From the discussion of the zero field response in the last section it is
evident that the time $t_q$ elapsed after the quench and before application of
the pump-field is a very important parameter.
For small $t_q$ some transient features are expected due to the interplay
of the approach of the aging-regime and the additional non-equilibrium features
induced by the application of the pump-field.
Of course, the cross-over to a $\Om$ independent behavior will depend crucially
on the pump-frequency $\Om$ as this determines the time $t_p$ of the imposed
non-equilibrium situation.
This is demonstrated in Fig.3a, where $\D\chi(\t)$ is plotted for $d\!=\!3$,
$T\!=\!0$, $t_w\!=\!0$, $\Om=0.1, 10$ and several values of $t_q$.
From this figure two features are evident immediately. First of all it is seen
that $\D\chi(\t)$ is non-zero only in a limited time interval and the time of
the maximum modification depends on the burn frequency $\Om$, thus
demonstrating dynamic heterogeneous behavior.
Additionally, the curves for $t_q\!=\!0$ differ from the others in that they
change from positive to negative values in a limited time range. This transient
behavior also depends on $\Om$.
However, it is always possible to choose $t_q$ in a way that the mentioned
interplay between the two sources of transient features can be neglected.
Thus, it is interesting to consider the asymptotic regime determined by
\[
t_q\gg 1\quad\mbox{where}\quad t_q=t_q(\Om).
\]
The detailed dependence of $t_q$ on $\Om$ has to be found empirically in the
sense that no transient effects should show up in the modified response for
a given pump frequency $\Om$.
In order to further demonstrate the relative independence of the results from
the chosen value of $t_q$, in Fig.3b I have plotted the value of the maximum
modification, $\D\chi_{max}=\D\chi(\t_{max})$ vs. $t_q$ for
$\Om=0.01, 0.1, 10$ in a scaled way.
This plot demonstrates the features already mentioned above.
For small $t_q$ there is some time interval in which $\D\chi(\t)$ is negative
and therefore the maximum is reduced. The small hump around
$t_q\sim 2\pi/\Om$ in the curves is roughly located at those values of $t_q$
where also the negative part vanishes.
Finally for long $t_q$ a plateau is observed and the results are independent
of $t_q$.
Obviously, the minimum value of $t_q$ satisfying the constraint that
$\D\chi(\t)$ is independent of $t_q$ depends on $\Om$.
In the whole range of $\Om$ considered in the present paper it turned out that
a value of $t_q\!=\!10^6$ is sufficient. Therefore all further calculations
are performed for $t_q\!=\!10^6$ unless stated otherwise.
(I have checked via explicit calculations that
$\chi(t_q+t_p,\t)-\chi(t_q,\t)\simeq 0$ independent of $\Om$. Otherwise, the
interpretation of the $\Om$-dependence of $\D\chi(\hat{t},\t)$ in terms of dynamic
heterogeneities would be meaningless.)
From eqns.(\ref{g.as.def}) and (\ref{del.W}) it is seen that in this regime
one has for $\D_0(t_q+\t)$ for arbitrary dimension ($t_m=$Min$(\t,t_p)$):
\be\label{del.as}
\D_0(t_q+\t)=(8\pi)^{-d/2}\int_0^{t_m}\!dt_1\int_0^{t_m}\!dt_2
			\sin{(\Om t_1)}\sin{(\Om t_2)}
			{ g\!\left(\t-\frac{t_1+t_2}{2}\right)\over
			[(t_q+t_1)(t_q+t_2)]^{d/4}}
\quad t_q\gg 1
\ee
This expression along with eq.(\ref{R.DR.allg}) for the modification of the
response also explains the observed relatively weak dependence of
$\D\chi(\t)$ on $t_q$. In lowest order
$\D_0(\hat{t}+\t')$ behaves as $t_q^{-d/2}$ and according to eq.(\ref{W.as})
the same holds for $W(\hat{t}+\t')$. Therefore, $R$ and $\D R$ are independent
of $t_q$ in this order and the same holds for $\D\chi$. An explicit
$t_q$-dependence enters only in higher order.

For $T=0$, the above expression is only needed for $t_m=t_p=2N\pi/\Om$, where
$N$ denotes the number of cycles of the sinusoidal field.
The reason for the fact that smaller values of $t_m$ are irrelevant in this
case is that according to eq.(\ref{W.del.T0}) $\D(t)^2=\D_0 (t)$ and that the
response is measured only after the pump-field is switched off.
Therefore, the question as to which extent the results depend on the number of
cycles naturally occurs.
Fig.3c shows $\D\chi(\t)$ as a function of $\t$ for $N=1, 5, 10$ for
$\Om=0.1$.
The curves for $N=5$ and $N=10$ are hardly distinguishable in this plot.
The saturation of the maximum amplitude is demonstrated in Fig.3d, where
$\D\chi_{max}$ is plotted vs. $N$ for $\Om=0.1$ and $10$ ($t_q=10^6$).
For $N$ values larger than roughly 6, $\D\chi_{max}$ become $N$-independent.
In all of these calculations one has $t_q\gg t_p$, whereas the opposite limit
is met at small $t_q$ as discussed above, although for $N\!=\!1$.
This demonstrates that the field sequence of NHB is unable to drive the system
much further away from equilibrium than it already is due to the quench
at $t=0$.

After this consideration of the influence of the parameters $t_q$ and $N$ now
the more important issue of the $\Om$-dependence of $\D\chi(\t)$ will be
discussed.
Before presenting the results of model calculations it is instructive to
investigate analytically the limiting behavior of $\D_0(t_q+\t)$ according to
eq.(\ref{del.as}) which in turn determines the behavior of $\D\chi(\t)$.
One finds that $\D_0$ vanishes in the limits of large and small $\Om$
according to
\Be\label{del.limits}
&&\Om\to 0:\quad\D_0(\hat{t}+\t)\sim\Om^{d/2-2}\quad(\Om t_q\gg 1)\nonumber\\
&&\hspace{1.9cm}\D_0(\hat{t}+\t)\sim\Om^{d-2}\hspace{0.7cm}(\Om t_q\ll 1) \\
&&\hspace{-0.25cm}\Om\to\infty:\quad\D_0(\hat{t}+\t)\sim\Om^{-4}
\quad\hspace{0.6cm}\mbox{$\forall$ d}\nonumber
\Ee
demonstrating that the modification induced by the NHB pulse sequence in
principle depends on $\Om$.
The universal $\Om^{-4}$-dependence for large $\Om$ can easily be understood
by slightly rewriting eq.(\ref{del.W}) in the form ($M\!\equiv\!\Om t_m$)
\[
\D_0(t_q+\t)=\Om^{-2}\int_0^{M}\!dt_1\int_0^{M}\!dt_2\sin{t_1}\sin{t_2}
	W\left(t_q+{t_1\over\Om}\right)	W\left(t_q+{t_2\over\Om}\right)
		g\left(\t-{t_1+t_2\over 2\Om}\right)
\]
The behavior for large $\Om$ is obtained via second-order expansions of
$W(t_q+\Om^{-1}x)$ and of $g(\t-\Om^{-1}x)$. In a simple calculation one then
finds the quoted $\Om^{-4}$ behavior.
Therefore, this represents a universal result which is valid for {\it any}
model obeying eq.(\ref{del.W}). As the $\Om$-dependence of $\D_0(t_q+t)$
uniquely determines the one of $\D(t_q+t)^2$, this result additionally holds
for all temperatures, including the disordered paramagnetic phase.

In Fig.4a the dependence of $\D\chi(\t)$ on $\Om$ is demonstrated in detail
for $d\!=\!3$.
It is clearly seen that $\D\chi(\t)$ shows a very pronounced $\Om$-dependence
for $\Om\!<\!0.1$ which, however, diminishes with increasing $\Om$.
Additionally, it is evident that the spectral modifications become 'broader',
i.e. are non-vanishing in a larger time interval, as $\Om$ decreases.

In Fig.4b I have plotted the time of the maximum modification, $\t_{max}$ vs.
$\Om$. From this plot one can see that $\t_{max}$ varies as $\Om^{-1}$ for
small $\Om$.
This is the behavior typical for an extremely broad distribution of relaxation
times.
Additionally, $\t_{max}$ becomes independent of $\Om$ for $\Om\sim 10$.
Thus, for the s-FM model in $d\!=\!3$, the response is dynamic homogeneous
in the short time regime, whereas it is dynamic heterogeneous for long times.
Therefore, the aging dynamics in this model is dynamic heterogeneous.
Also, it is seen from Fig.4b that the behavior is independent of the time
$t_q$ elapsed after the quench.

Another important question regards the time scale of the 'recovery' of the
modification, i.e. the waiting time dependence. In ref.\cite{dieze01} I have
found that in a model of reorientational dynamics there is no extra time scale
for the recovery. However, in the experiments on the relaxor materials a very
long recovery time scale has been found\cite{KSB98}. In order to investigate
this question, in Fig.5a I have plotted the maximum modification (normalized to
the value at $t_w\!=\!0$), $\D\chi(\t_{max})_{norm.}$, vs. scaled waiting
time $\Om t_w$ for a variety of burn frequencies $\Om$. It is evident that
the life-time strongly depends on $\Om$ for small $\Om$ and that this
dependence diminishes for larger $\Om$.
It has to be mentioned at this point that the form of the modifications
hardly change for different waiting times.
In order to have a simple measure for the recovery times, I have fitted the
curves of Fig.5a to a Kohlrausch function, $\exp{\{-(t_w/\t_K)^{\b_K}\}}$,
and plotted the resulting time scales $\t_K$ vs. $\Om$ in Fig.5b.
I included the time of the maximum modification for $t_w\!=\!0$ (cf. Fig.4b)
in this plot (dot-dashed line).
As the behavior of these two time scales is almost identical to within a
factor of two, the conclusion to be drawn from these calculations is that
there does not exist an extra time scale for the recovery in the s-FM model.
This appears plausible on intuitive grounds because the only time scale set
by the pump is roughly $1/(2\pi\Om)$ in the domain growth regime and
just unity ($1/(2\pi))$ in the short-time regime.
This is qualitatively different from a complex domain-structured system like
the relaxor materials, where pinning effects play a dominant role.
\subsubsection*{T$>$0:}
So far, I have considered $T\!=\!0$ solely.
As already mentioned in the last section, the response in the ferromagnetic
phase is independent of temperature. For the NHB field sequence, however,
eqns.(\ref{del.W}) and (\ref{R.DR.allg}) show that $\D R$ and correspondingly
also $\D\chi(\t)$ do depend on temperature.
The physical reason for this dependence is quite clear.
Though the thermal noise is irrelevant for the linear response, this changes
under NHB conditions due to the aligning effect of the pump-field.
Here, thermal fluctuations tend to counterbalance the induced alignment.

In order to investigate the importance of this effect, eq.(\ref{del.W})
is solved numerically for finite temperatures for $t_q\!\gg\!1$, i.e. using
eq.(\ref{W.as}) in the expression for $\D_0(t)$. (To ensure that the physical
solution of the equation is met, I first performed calculations for very small
T, in which case eq.(\ref{del.W}) can be solved in terms of an expansion in
$T/T_c$.) The results of such temperature dependent calculations
are shown in Fig.6a, where $\D\chi(\t)$ is plotted for various $\Om$ and for
temperatures up to $0.5T_c$ using $t_q\!=\!10^{6}$ and $t_w\!=\!0$.
Two features are evident by inspection of that plot.
First, the position of the maximum modification hardly changes as a function
of temperature. It should be mentioned that also the shape of the $\D\chi(\t)$
does not change as a function of temperature. This means that no effect
of 'motional narrowing' is observable.

The most prominent feature is the increasing intensity of $\D\chi(\t)$. In
order to demonstrate this behavior in more detail, the intensity at the time
of the maximum modification, $\D\chi(\t_{max})$, is shown as a function of
reduced temperature $1-T/T_c$ in Fig.6b for $\Om\!=\!0.01, 100$.
This plot demonstrates a scaling behavior of these quantities.
The exponents, however, are different for various frequencies and do not seem
to have an obvious explanation.
It would be interesting to further analyze the $\Om$-dependence of
$\D\chi(\t_{max})$, which, however, is beyond the scope of the present study.

The conclusion to be drawn from these calculations is that the effect of
temperature is seen mainly in the amplitude of the modification, but narrowing
effects are absent. This means that the nature of the dynamic heterogeneities
are not affected by temperature-effects. Therefore, in the following, I will
concentrate on $T=0$ for simplicity.
\subsection*{B. Spherical-ferromagnet: NHB for varying dimension}
In contrast to the s-SK model the s-FM model offers the opportunity to study
the behavior of the observed dynamic heterogeneities as a function of spatial
dimension.
As already mentioned in Sect.II the fluctuation-dissipation ratio $X_\infty$
vanishes for the s-FM model independent of spatial dimension. Additionally, it
is known\cite{GL00} that the usual static critical exponents take on their
mean-field values for $d\!>\!4$ while some dynamical exponents depend on
dimension for arbitrary $d$.
The only hint for a change in the dynamic properties of coarsening models
for $d\!>\!4$ stems from the large N-model, where it has been shown that the
aging contribution to the integrated response changes qualitatively for
$d\!>\!4$\cite{CLZ02}.
The two-time quantities in the spherical models considered here, however, do
not show any signature of a change in behavior around $d\!=\!4$.
Therefore, in this subsection, the dependence of the behavior of $\D\chi(\t)$
and thus of the dynamic heterogeneities on spatial dimension will be
investigated for $T\!=\!0$.

Fig.7a shows a plot of $\D\chi(\t)$ versus $\t$ for $d\!=\!5$ (upper panel) and
$d\!=\!7$ (lower panel), which is to be compared to Fig.4a.
(I do not consider even spatial dimensions like $d\!=\!6$, because the
calculations are much more involved for technical reasons.)
In both cases the $\Om$-dependence is much weaker than in $d\!=\!3$.
Whereas there is some weak $\Om$-dependence for $d\!=\!5$, such a dependence
is hardly visible for $d\!=\!7$.
To quantify this diminishing $\Om$-dependence, in Fig.7b I have plotted the
time, $\t_{max}$, of the maximum modification as a function of $\Om$.
Whereas one has a $\Om^{-1}$ dependence in the long-time regime for
$d\!=\!3$, this is weaker, roughly $\Om^{-1/2}$, for $d\!=\!5$ and finally in
$d\!=\!7$ and $d\!=\!9$ there is no visible $\Om$-dependence. This means that
the dynamics in $d\!=\!7$ an $d\!=\!9$ is {\it dynamic homogeneous} in the
short-time {\it and} the long-time regime.
Thus, also the aging dynamics becomes dynamic homogeneous for higher spatial
dimension.

The conclusion from these calculations is that the dynamics of the s-FM model
becomes dynamic homogeneous in the mean-field limit, $d\!\gg\!1$.
However, as pointed out earlier, the mean-field limit holds for all $d\!>\!4$
when concerned with static properties. For $d\!=\!5$, Fig.7 reveals that the
aging dynamics still is heterogeneous, although the shape of an effective
distribution of relaxation times is changed relative to $d\!=\!3$.

\subsection*{C. NHB in the spherical Sherrington-Kirkpatrick model}
As already noted in the preceeding section the dynamic behavior of the s-FM
model in $d\!=\!3$ and the s-SK model is very similar in the low temperature
phase.
Concerning the modified response $\chi^*(\t)$ the same holds true, in
particular in the limit of large $t_q$.
The only difference stems from the functions $g(t)$ occuring in the expressions
for $\D_0(t_q+\t)$ and $R(\hat t+\t,\hat t +\t')$,
eqns.(\ref{R.DR.allg},\ref{del.as}), for the two models.
Plotting the two functions $g_{FM}(t)$ and $g_{SK}(t)$ reveals that
$g_{FM}(t)\!\simeq\! g_{SK}(t/3)$ in a good approximation for moderate $t$ and
asymptotically one has
$g_{SK}(t)\simeq[(8\pi)^{3/2}/(32\pi)^{1/2}]g_{FM}(t)\!\simeq\!12.57g_{FM}(t)$,
cf. eq.(\ref{g.as.def}).
For large $\Om$ small times $\t$ are relevant and therefore one expects
$\D\chi_{SK}(\t)\simeq\D\chi_{FM}(\t)$ to hold with the difference that
the maximum modification occurs at $\t_{max}^{SK}\!\simeq\!3\t_{max}^{FM}$.
On the other hand, for small $\Om$ long times are most important and all
functions $g(t)$ can be replaced by their asymptotic values. Thus, one
roughly has $\D\chi_{SK}(\t)\simeq[(8\pi)^{3}/(32\pi)]\D\chi_{FM}(\t)$
and $\t_{max}^{SK}\!\simeq\!\t_{max}^{FM}$.
In Fig.8a I have plotted $\D\chi(\t)$ for both models for $\Om=10^{-4}$
(upper panel) and $\Om=100$ (lower panel).
Apart from a factor of two in the amplitude for the larger frequency the
behavior just discussed is recovered.

Fig.8b shows the time of the maximum modification, $\t_{max}$ versus
frequency $\Om$.
The full line is for the 3d s-FM, cf. Fig.4b,  and the dashed line is for the
s-SK model.
Also included is $(\t_{max}^{SK}/3)$ for $\Om>1$ (dot-dashed line).
From the above discussion it is clear that one just finds the expected
behavior: for very small $\Om$ one has $\t_{max}^{SK}\!\simeq\!\t_{max}^{FM}$
whereas $\t_{max}^{SK}\!\simeq\!3\t_{max}^{FM}$ holds for large $\Om$.

These considerations show that the qualitative behavior of the s-FM model for
$d\!=\!3$ and the s-SK model are extremely similar in the low temperature
phase, including the dynamic heterogeneities.
The similarity between the 3d s-FM and the s-SK model is well known and it has
been argued earlier that the s-SK model does not generically behave like a
spin glas, concerning both, the statics and the dynamics\cite{ZKH00}.

The fact that there are dynamic heterogeneities observable in the s-SK model
might appear somewhat astonishing in view of the fact that the s-SK
model is a mean-field model.
However, this result is not unique.
The first observation of dynamic heterogeneous relaxation in a mean-field
model was reported in ref.\cite{leticia01}.
Also the Monte Carlo simulations on the Sherrington-Kirkpatrick model
mentioned above\cite{let.reply} show dynamic heterogeneous behavior.
The present results therefore confirm the fact that an identification of
{\it dynamic} and {\it spatial} heterogeneities is not possible in general.
\section*{IV. Conclusions}
In the present paper I presented calculations of the response of simple
spherical spin models to the field sequence of the nonresonant hole burning
experiment, a technique allowing the detection of dynamic heterogeneities if
they exist.
The calculations were restricted to the generic situation of aging in the
low temperature phase. It was assumed that the system is quenched from
infinite temperature to $T\!<\!T_c$ at the beginning, correlations in the
initial conditions thus being neglected completely. The equilibrium dynamics
will be considered in a forthcoming publication.

The main result of the present paper is the fact that the non-equilibrium
dynamics of the s-FM model in $d\!=\!3$ is dynamic heterogeneous in the
aging regime whereas it appears homogeneous in the short-time regime.
In higher dimensions, $d\!\geq\!7$, the response becomes dynamic homogeneous
also in the aging regime.
Therefore, the nature of the aging dynamics changes and in the mean-field
limit, $d\!\gg\!1$, the model displays homogeneous dynamics.
This is what one would expect as in this limit each spin interacts with an
infinite number of neighboring spins.
However, for the s-SK model, a simple mean-field spin glass model, a behavior
very similar to that of the s-FM model in $d\!=\!3$ is found.
This explicitly demonstrates that the existence of dynamic heterogenities does
not tell us anything about spatial heterogeneities in the system.
Furthermore, a strongly heterogeneous dynamics has also been observed in the
aging behavior of a short range spin glass model\cite{CCCK02} showing strong
relations between local responses and correlations.
The mere existence of dynamic heterogeneities in a short range spin glass
is of course to be expected and here also the spatial aspects of these
heterogeneities are of importance.

When considering the spherical two-spin interaction models, it is tempting to
associate some disorder with the spherical constraint which forces the equal
time correlation function to unity at all times.
The lengths of the spins are not static quantities and are random to some
extent.
This might be viewed as a kind of 'dynamic' disorder and the similarities
between the s-FM model in $d\!=\!3$ and the s-SK model hint towards the
irrelevance of the quenched disorder in the latter when concerned with
dynamic quantities.
On a speculative level the fact that the dynamics in the s-FM model becomes
homogeneous for $d\!\gg\!1$ can be understood from the following argument.
Assuming the existence of an effective distribution of relaxation rates at a
given instant of time, i.e. for a given distribution of $s_i(t)^2$
the width of this distribution is expected to decrease with an increasing
number of nearest neighbor interactions and consequently the dynamics
becomes more homogeneous. This of course does not mean that the response
decays in an exponential way for large $d$ because the considered distribution
is not a static quantity.
Notice that this argument also implies that the life-time of the
heterogeneities is finite. Unfortunately, NHB does not allow to determine
this life-time.
It is, however, important to point out that the argument concerning the
increasing number of neighbors, though appealing with respect to the
s-FM model, has important drawbacks. Taking it serious in case of the
s-SK model would predict homogeneous dynamics in contrast to what is
observed.

Although the s-FM model is a typical model for phase-ordering kinetics,
there is no obvious relationship between the observed dynamic heterogeneities
and the domain size distribution in a coarsening system.
A possible way to investigate such a relationship could be to perform
calculations on the Ising model using spatially varying magnetic fields.
It would also be interesting to perform an analysis along the lines of
refs.\cite{CCCK02, MT03} in order to see whether the dynamic heterogeneities
in a coarsening system behave similar to those observed in glassy systems.
Furthermore, such calculations would allow to investigate the dependence on
spatial dimension and therefore to check whether the one observed in the
present paper for the s-FM model also is found for other domain-coarsening
models.

In summary, I have shown that heterogeneous aging can be observed in the
low-temperature phase of the spherical model of a ferromagnet. The aging
dynamics becomes homogeneous on increase of the spatial dimension despite the
fact that no qualitative change in the two-time quantities like the correlation
function or the response is observed.
Quenched disorder does not play any significant role with respect to
heterogeneities in spherical models.
\subsection*{Acknowledgements}
I thank R. B\"ohmer, G. Hinze, U. H\"aberle and R. Schilling for fruitful
discussions.
This work was supported by the DFG under contract Di693/1-1.

\newpage
\section*{Figure captions}
\begin{description}
\item[Fig.1 : ] The response function $R(\t+t_w,t_w)$ for the s-FM model
in the ferromagnetic phase, $T<T_c$, for $t_w=10^2,10^3,10^4,10^5$ (from upper
to lowest line). Upper panel: $d=3$, middle panel: $d=5$, lower panel: $d=7$.
\item[Fig.2 : ] The field sequence for the nonresonant hole burning (NHB)
experiment: A time $t_q$ after a quench from $T=\infty$ one or more cycles
of a strong sinusoidal field $h_p(t)=h_p\sin{(\Om(t-t_q))}$ are applied.
After a waiting time $t_w$ the response to an infinitesimally small field is
monitored.
\item[Fig.3 : ]
{\bf a:} The modification of the response, $\D\chi(\t)$ vs. $\t$ for
the s-FM, $d\!=\!3$. Upper panel: $\Om\!=\! 0.1$; lower panel:
$\Om\!=\! 10$.
The time after the quench from infinite temperature, $t_q$, is chosen as
$t_q\!=\! 0, 10, 10^6$.\\
{\bf b:} $\D\chi_{max, sc.}=\D\chi_{max}(t_q)/\D\chi_{max}(t_q\!=\!10^{-6})$
versus $t_q$ for $\Om=0.01, 0.1, 10$.\\
{\bf c:} $\D\chi(\t)$ versus $\t$ for $\Om=0.1$ and $t_q=10^6$ for
$N=1$ (full line), $N=5$ (dashed line) and $N=10$ (dotted line).
The curves for $N=5$ and $N=10$ are hardly to distinguish.\\
{\bf d:} $\D\chi_{max, sc.}=\D\chi_{max}(N)/\D\chi_{max}(N\!=\!1)$ as a
function of the number of cycles of the sinusoidal field, $N$ ($t_q=10^6$).
\item[Fig.4 : ]
{\bf a:} $\D\chi(\t)$ vs. $\t$ for $t_q\!=\!10^6$, $t_w\!=\!0$,
$d\!=\!3$ and $\log{(\Om)}=1, 0, -1, -2, -3, -4$ (from left to right).
\\
{\bf b:} Full line: $\t_{max}$ vs. $\Om$, dashed line: $\Om\t_{max}$ vs. $\Om$
for $t_q\!=\!10^6$, $t_w\!=\!0$. Additionally shown as the dot-dashed line is
$\t_{max}$ vs. $\Om$ for $t_q\!=\!1$ and $t_w\!=\!0$, demonstrating that the
behavior does not depende on $t_q$.
\item[Fig.5 : ]
{\bf a:} $\D\chi(\t_{max})_{norm.}$ vs. the scaled waiting time $\Om t_w$
for $\Om=10^{-4}, 1, 5, 10, 25, 50, 10^2$ (from left to right).\\
{\bf b:} Characteristic decay time $\t_K$ of a Kohlrausch-fit of the from
$\D\chi(\t_{max})_{norm.}=\exp{\{-(t_w/\t_K)^{\b_K}\}}$ ($t_q\!=\!10^6$).
The stretching parameter is approximately constant, $\b_K\!\simeq\!0.9$.
The dot-dashed line represents $\t_{max}$ for $t_w=0$ for comparison, cf.
Fig.4b.
\item[Fig.6 : ]
{\bf a:}
$\D\chi(\t)$ vs. temperature for $T=0,0.1,0.2,0.3,0.4,0.5$ for the burn
frequencies given in the respective panels. The remaining parameters are
$t_q\!=\!10^6$, $t_w\!=\!0$ and $d\!=\!3$.\\
{\bf b:}
The value of the maximum modification, $\D\chi(\t_{max})$,
vs. $1-T/T_c$. The other parameters are the same as in {\bf a}.
\item[Fig.7 : ]
{\bf a:}
$\D\chi(\t)$ vs. $\t$ for burn frequencies $\Om=1,10,20,50,100$ for
$d\!=\!5$ (upper panel) and $d\!=\!7$ (lower panel).
The remaining parameters are $t_q\!=\!10^6$ and $t_w\!=\!0$.\\
{\bf b:}
The time of the maximum modification, $\t_{max}$, for $d\!=\!3$, $d\!=\!5$
and $d\!=\!7$ vs. $\Om$. In the upper panel $\t_{max}$ is shown and in the
lower one the product $\Om\t_{max}$.
Full lines are for $t_q=10^{6}$ and the dot-dashed lines for $t_q=0$.
\item[Fig.8 : ]
{\bf a:}
$\D\chi(\t)$ vs. $\t$ for the 3-dimensional s-FM (full line) and the s-SK
model (dashed line). Upper panel: $\Om\!=\!10^{-4}$ and the factors
muliplying $\D\chi(\t)$ are $N(SK)=32\pi$ and $N(FM)=(8\pi)^3$. Lower panel:
$\Om\!=\!10^2$.
The remaining parameters are $t_q\!=\!10^6$ and $t_w\!=\!0$.\\
{\bf b:}
The time of the maximum modification, $\t_{max}$, for the 3d s-FM model (c.f.
Fig.4b, full line) and the s-SK model (dashed line) vs. $\Om$ for the same
parameters as in {\bf a}.
Also included is $\t_{max}^{SK}/3$ (dot-dashed line) for $\Om\!>\!1$.
\end{description}
\newpage
\begin{figure}
\includegraphics[width=15cm]{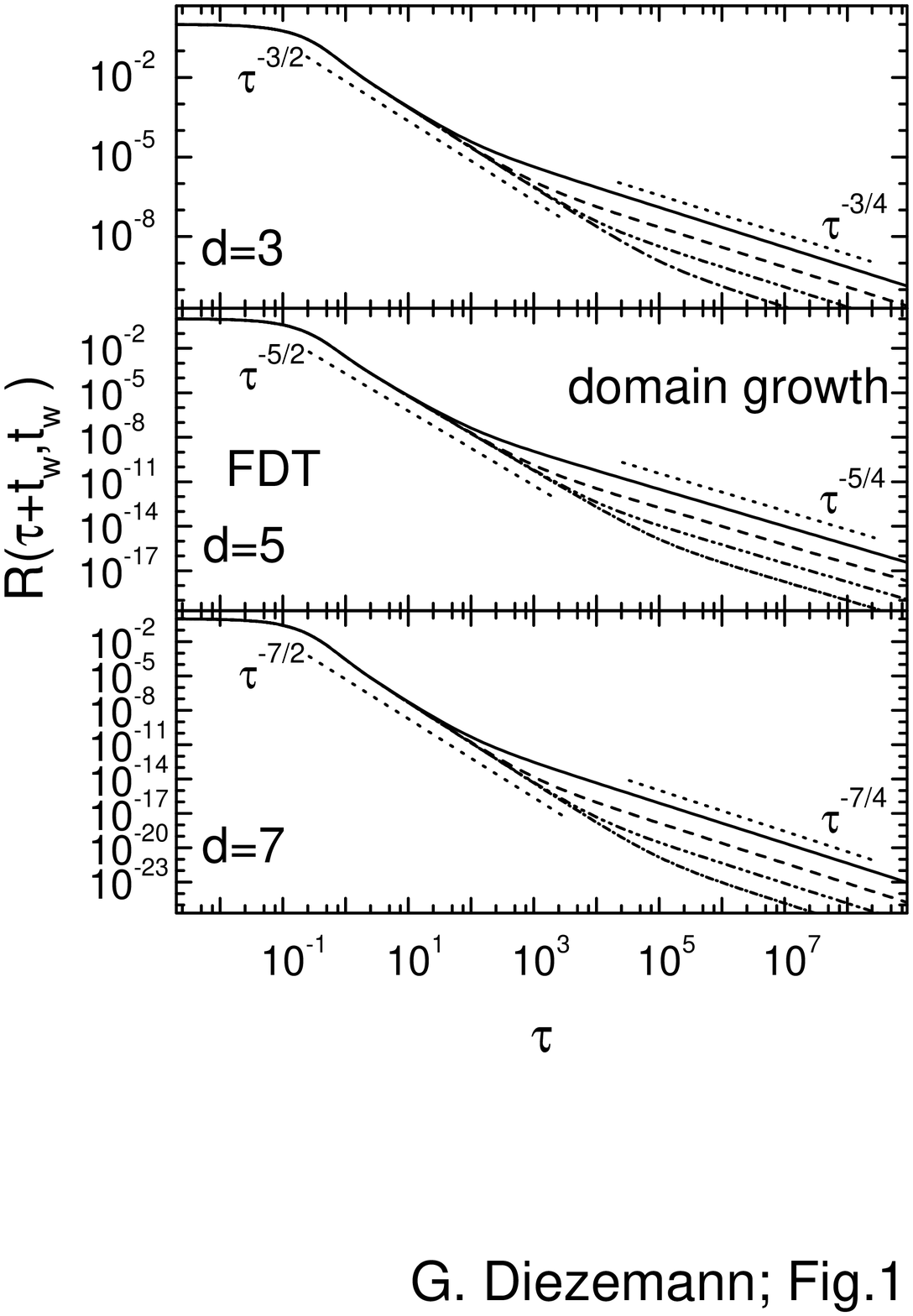}
\end{figure}
\newpage
\begin{figure}
\includegraphics[width=15cm]{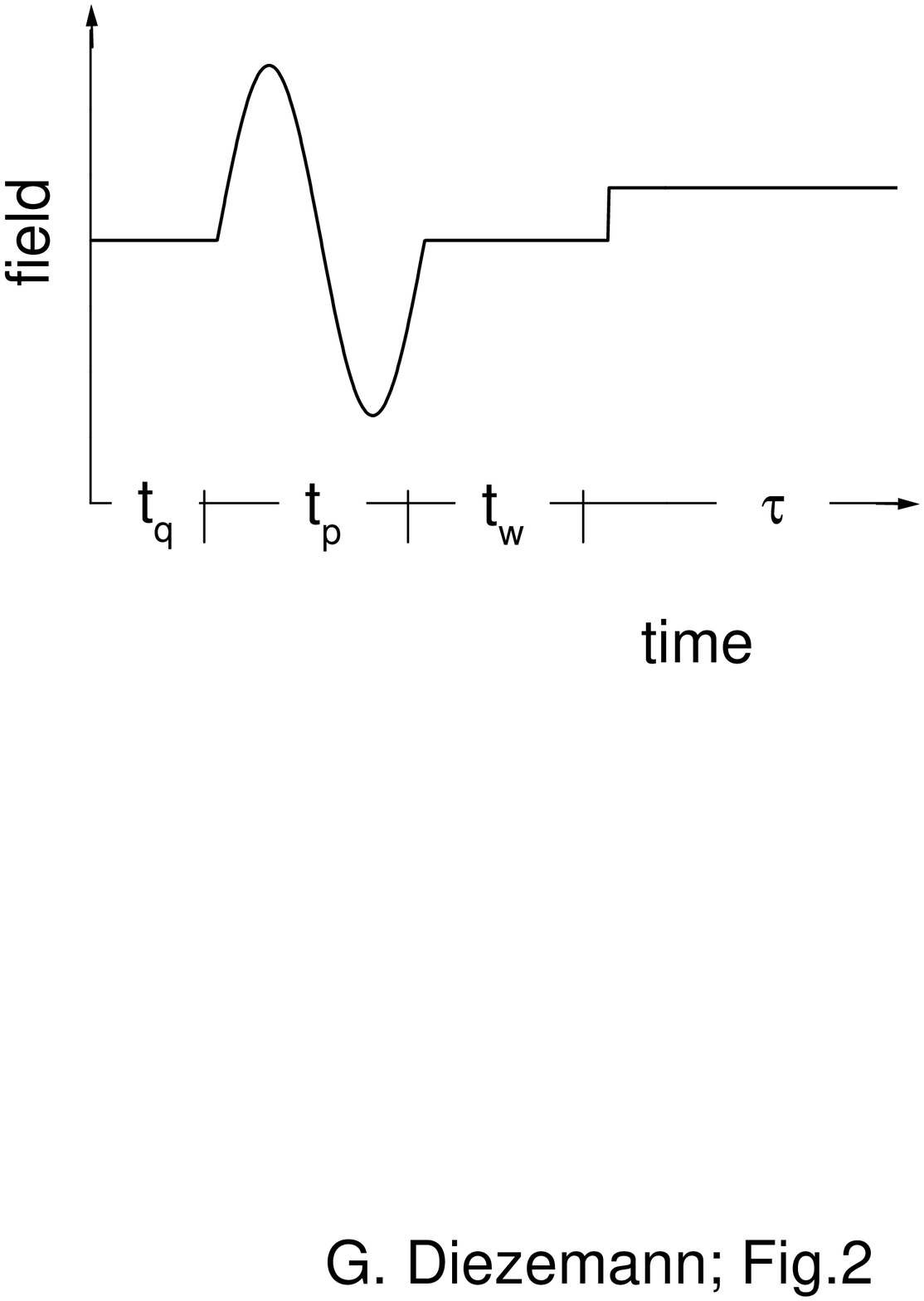}
\end{figure}
\newpage
\begin{figure}
\includegraphics[width=15cm]{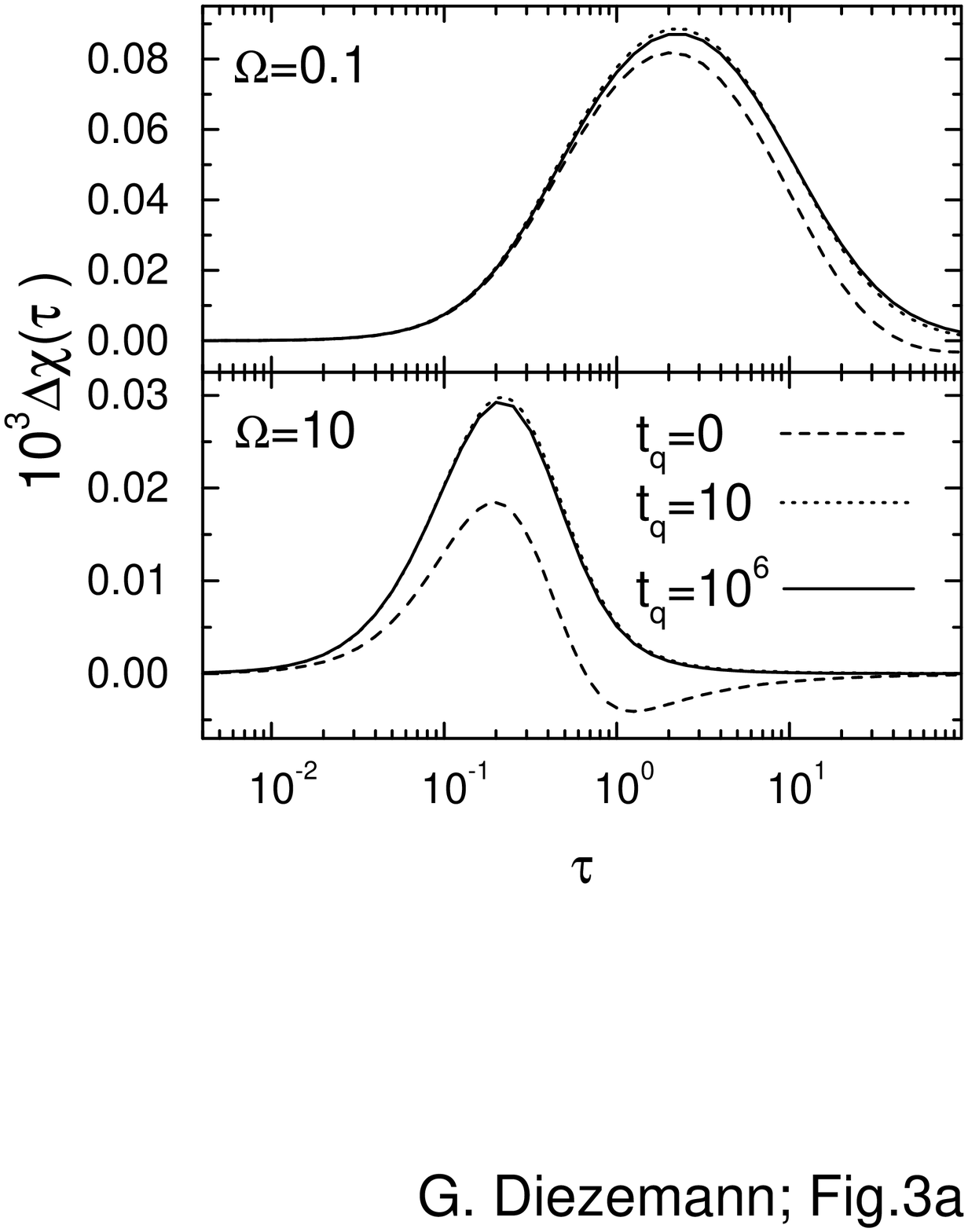}
\end{figure}
\newpage
\begin{figure}
\includegraphics[width=15cm]{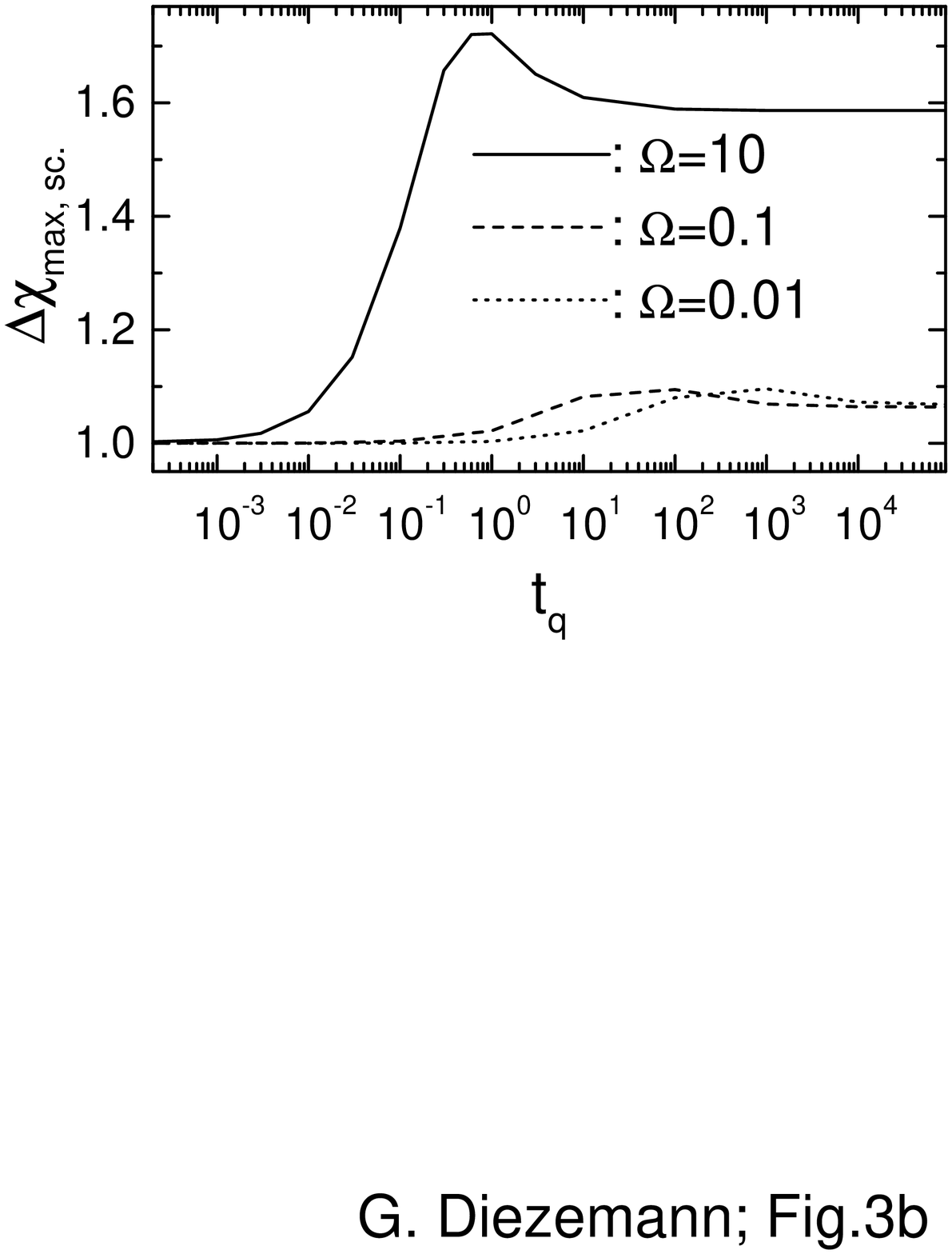}
\end{figure}
\newpage
\begin{figure}
\includegraphics[width=15cm]{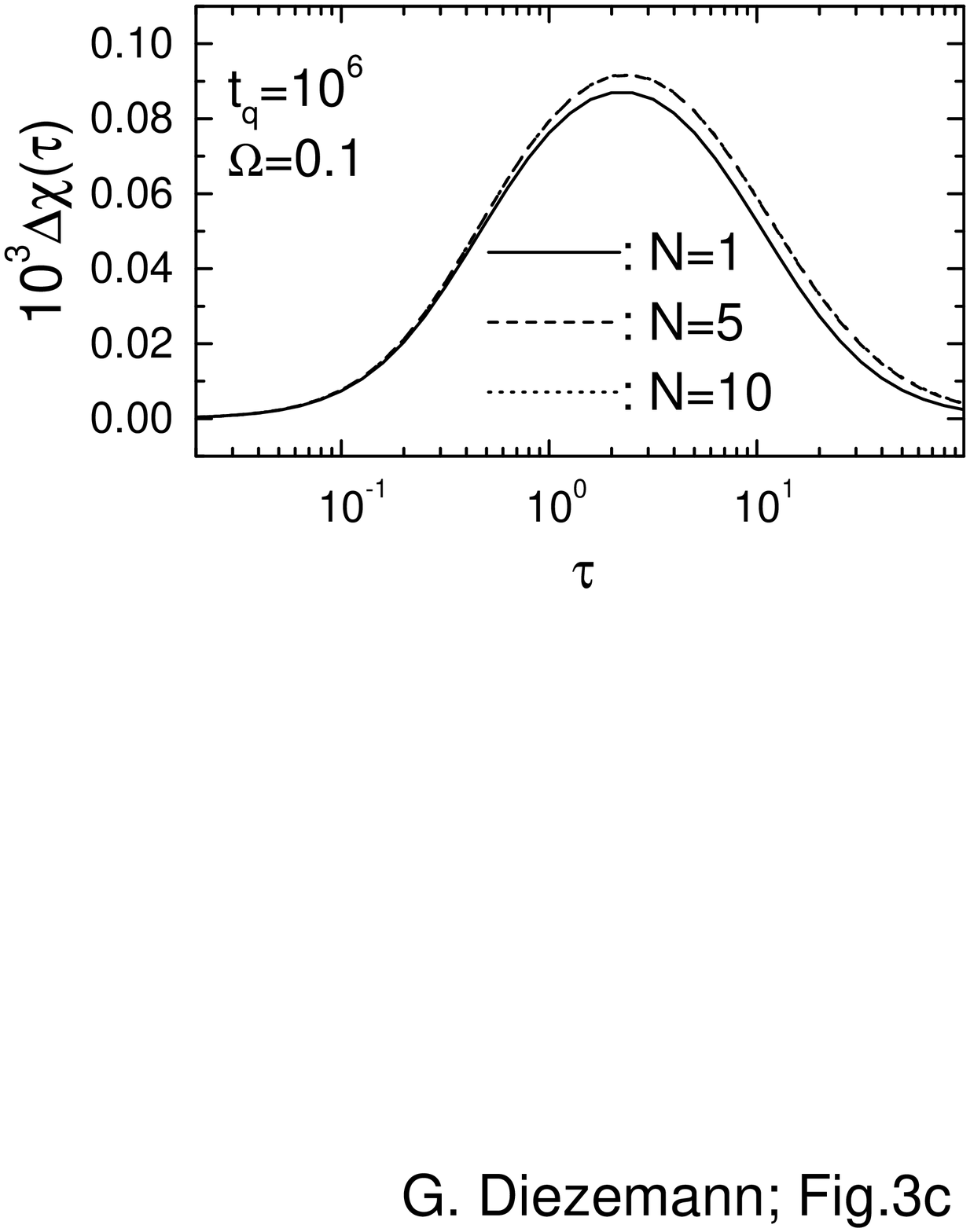}
\end{figure}
\newpage
\begin{figure}
\includegraphics[width=15cm]{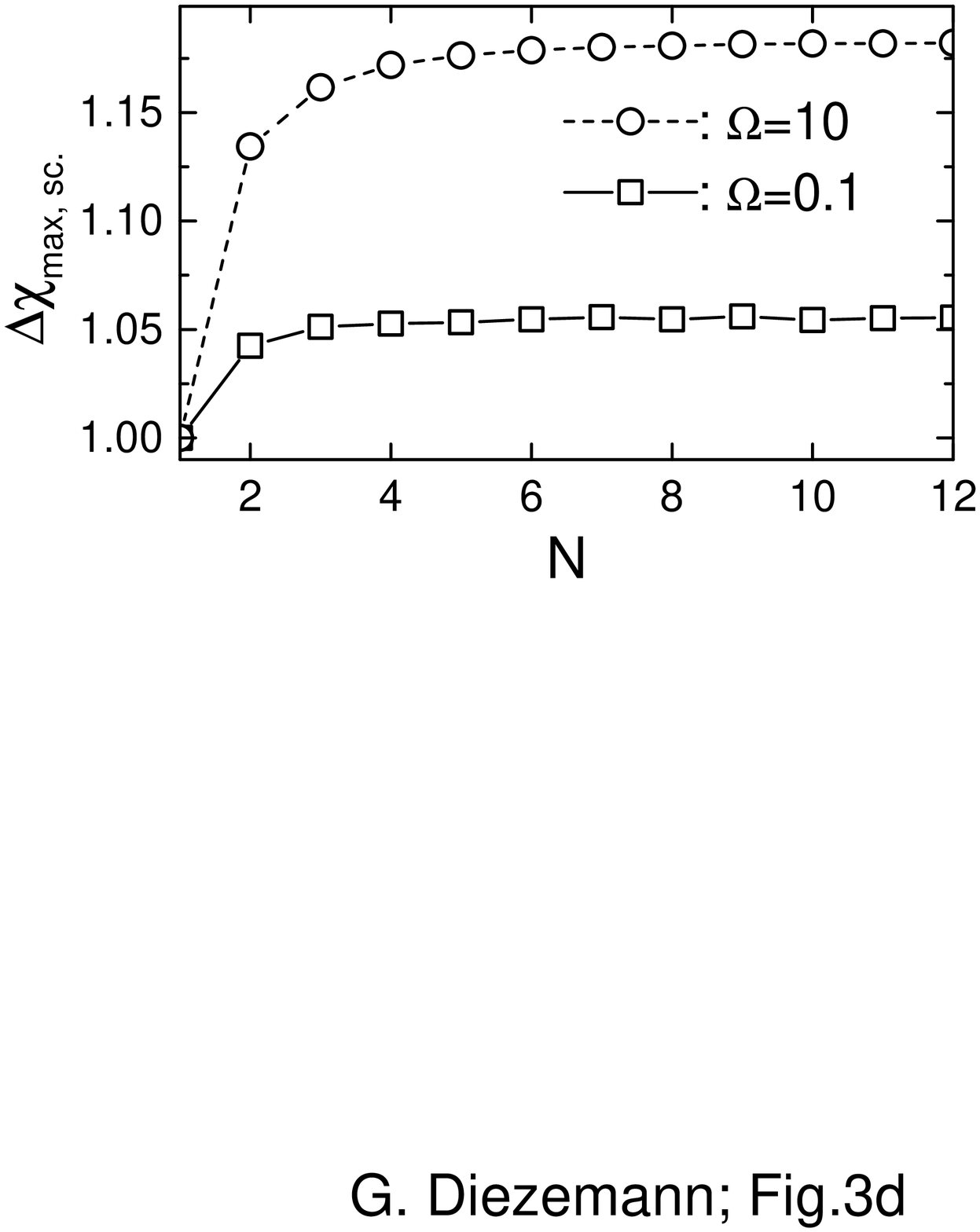}
\end{figure}
\newpage
\begin{figure}
\includegraphics[width=15cm]{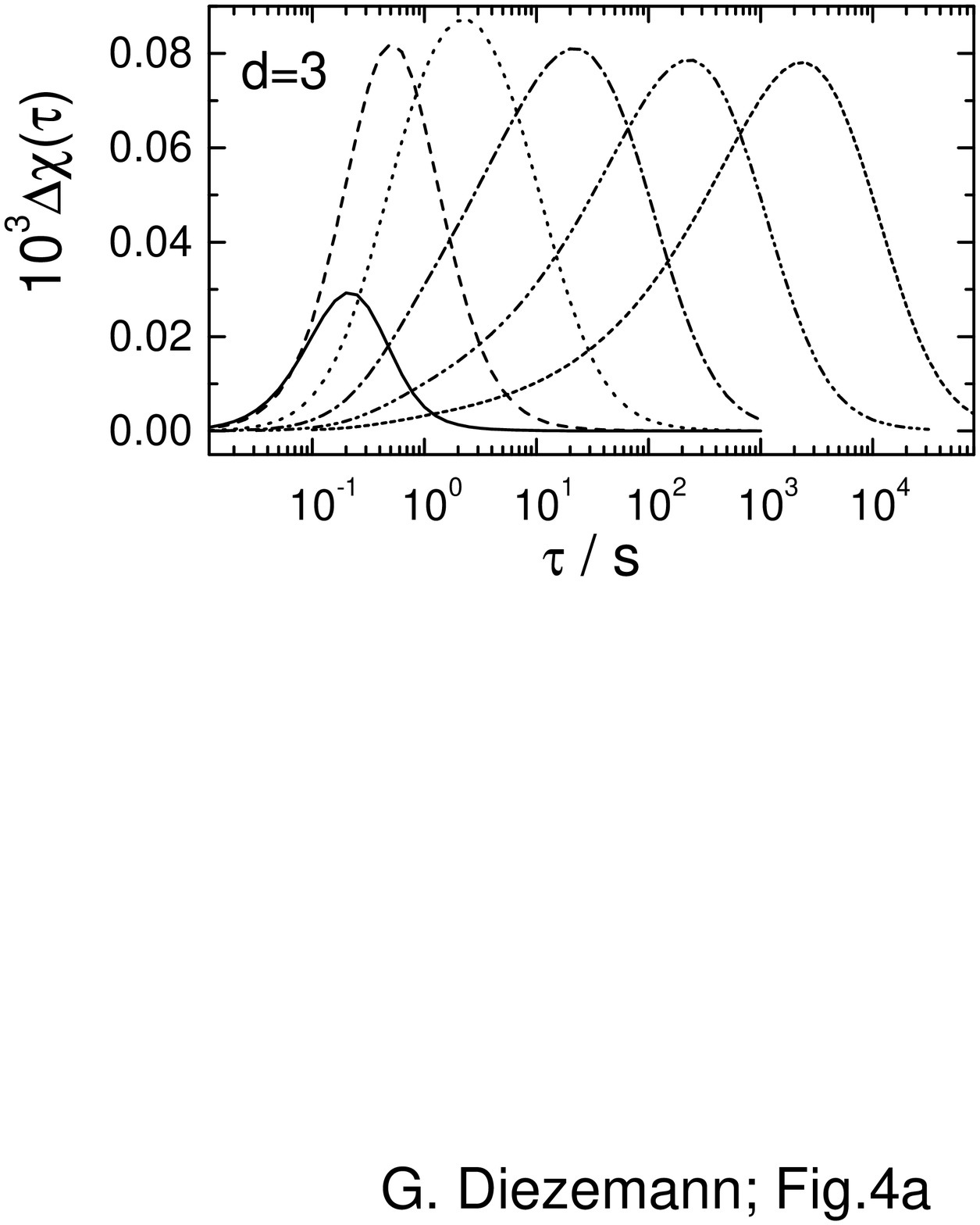}
\end{figure}
\newpage
\begin{figure}
\includegraphics[width=15cm]{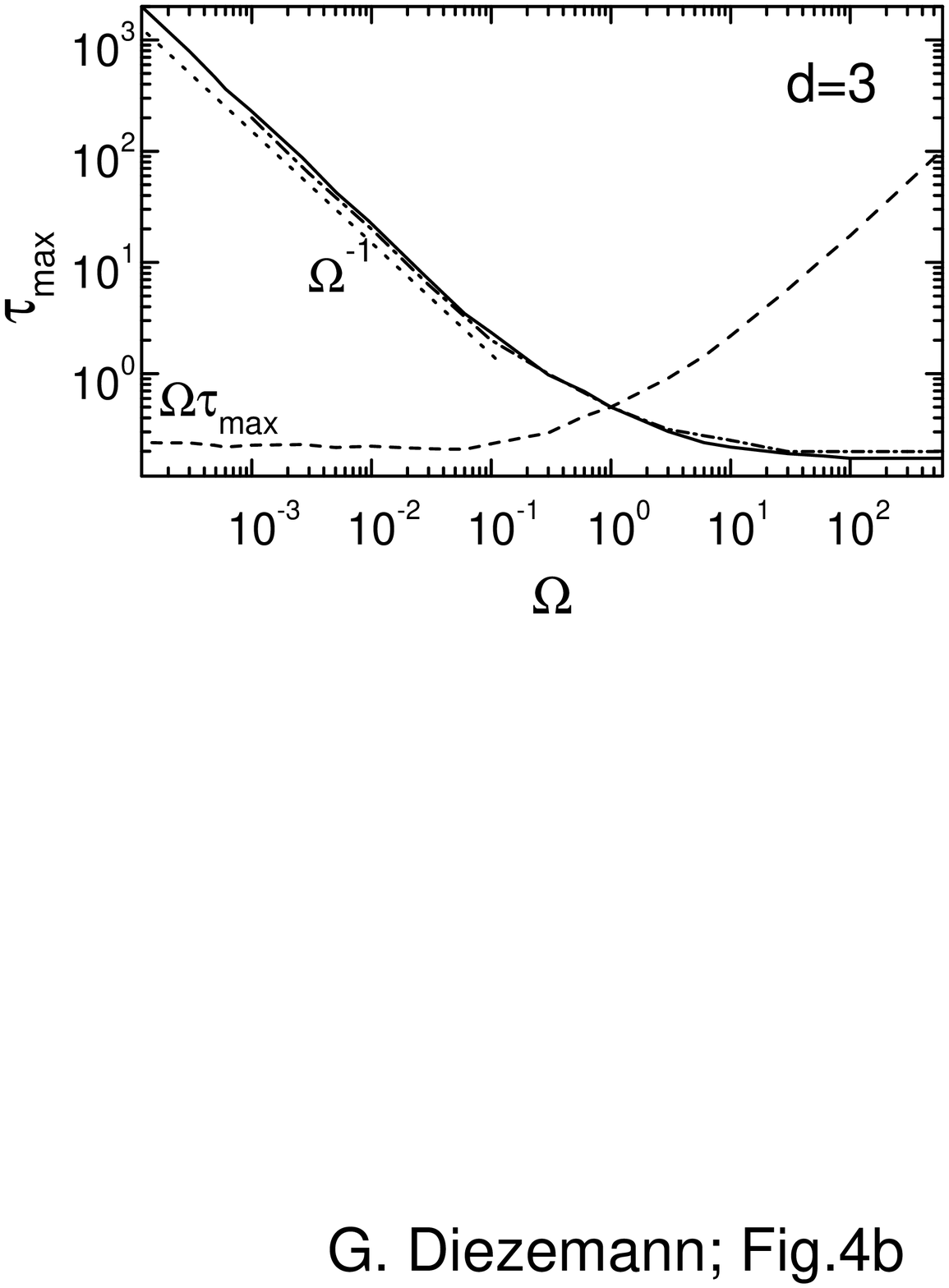}
\end{figure}
\newpage
\begin{figure}
\includegraphics[width=15cm]{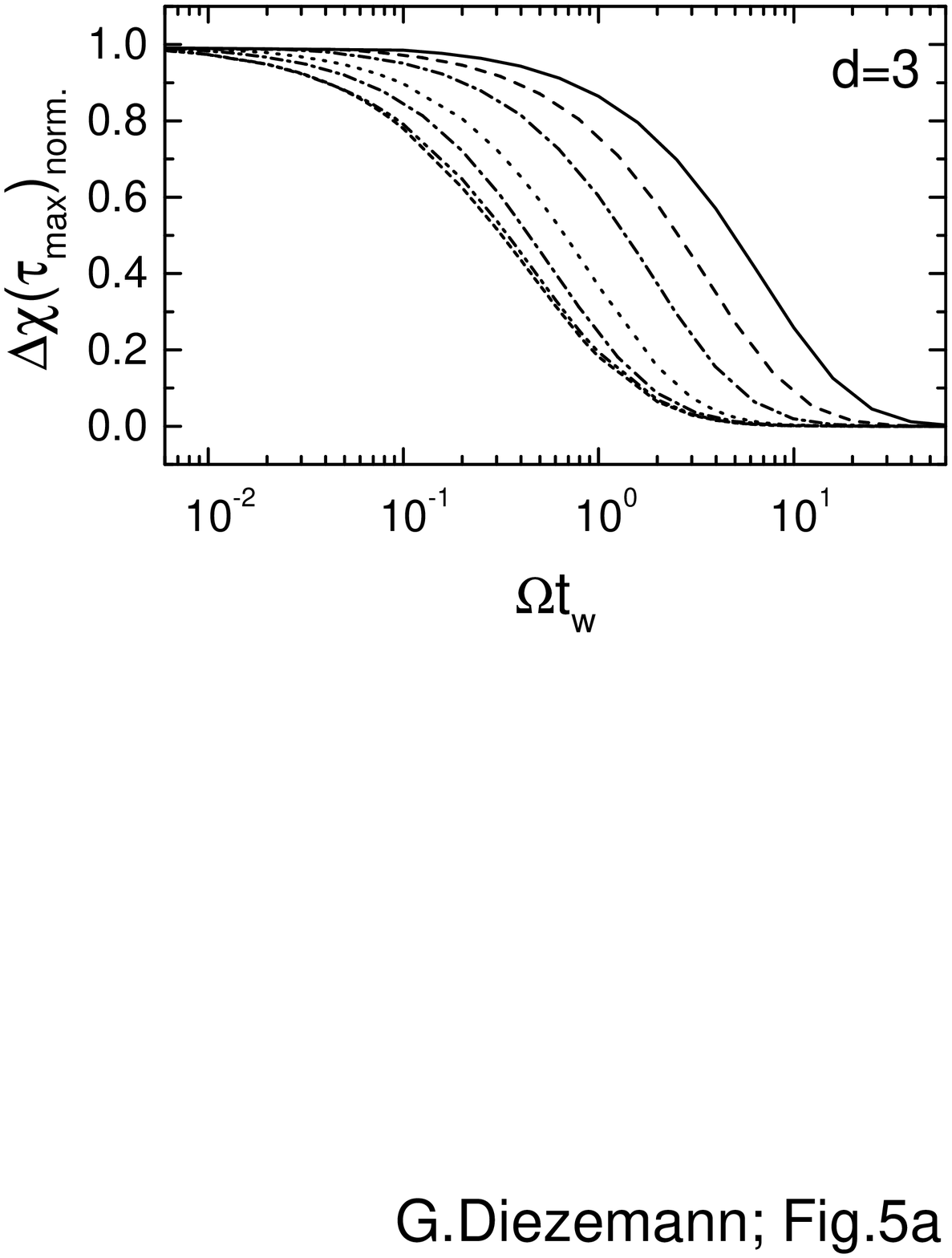}
\end{figure}
\newpage
\begin{figure}
\includegraphics[width=15cm]{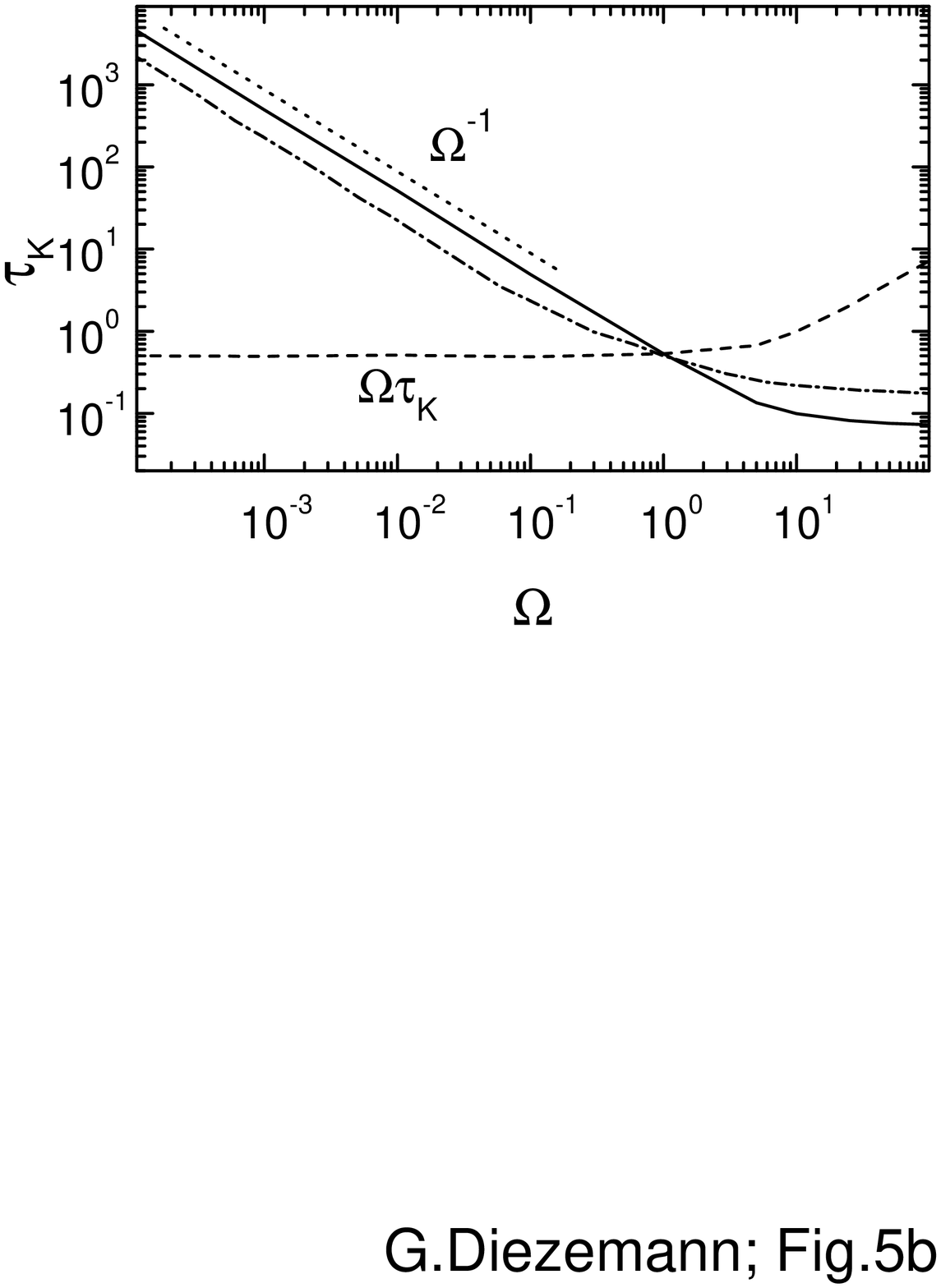}
\end{figure}
\newpage
\begin{figure}
\includegraphics[width=15cm]{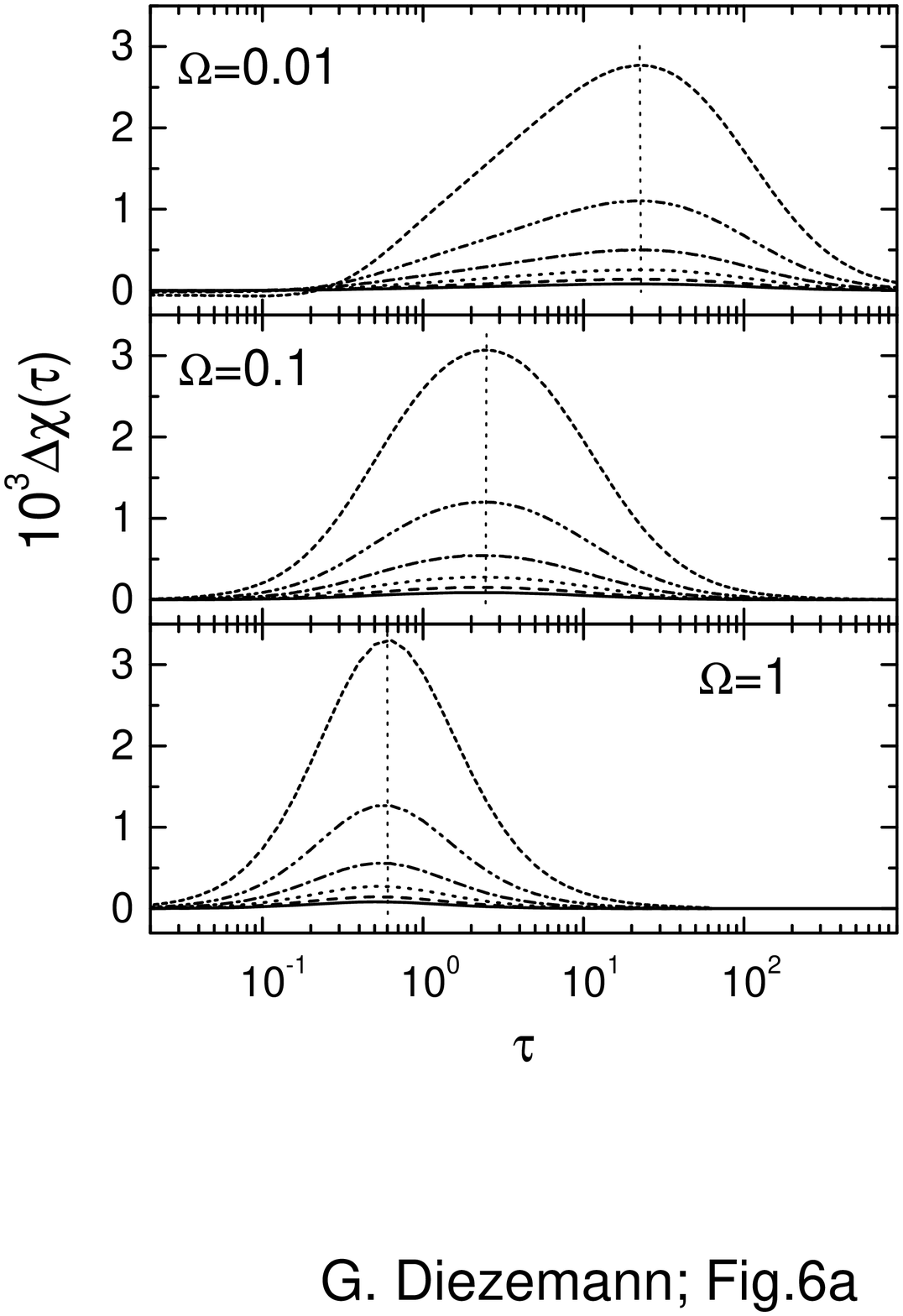}
\end{figure}
\newpage
\begin{figure}
\includegraphics[width=15cm]{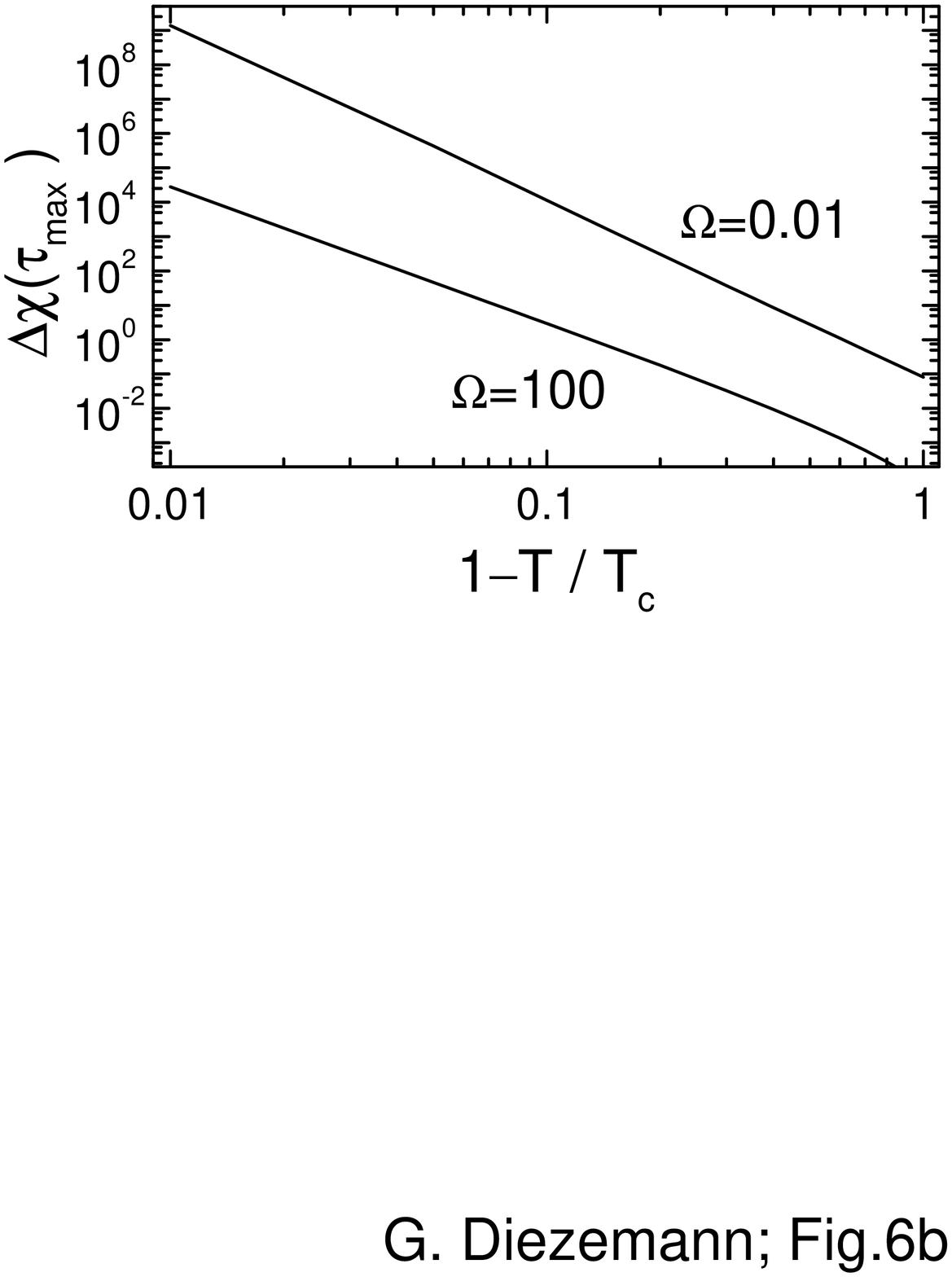}
\end{figure}
\newpage
\begin{figure}
\includegraphics[width=15cm]{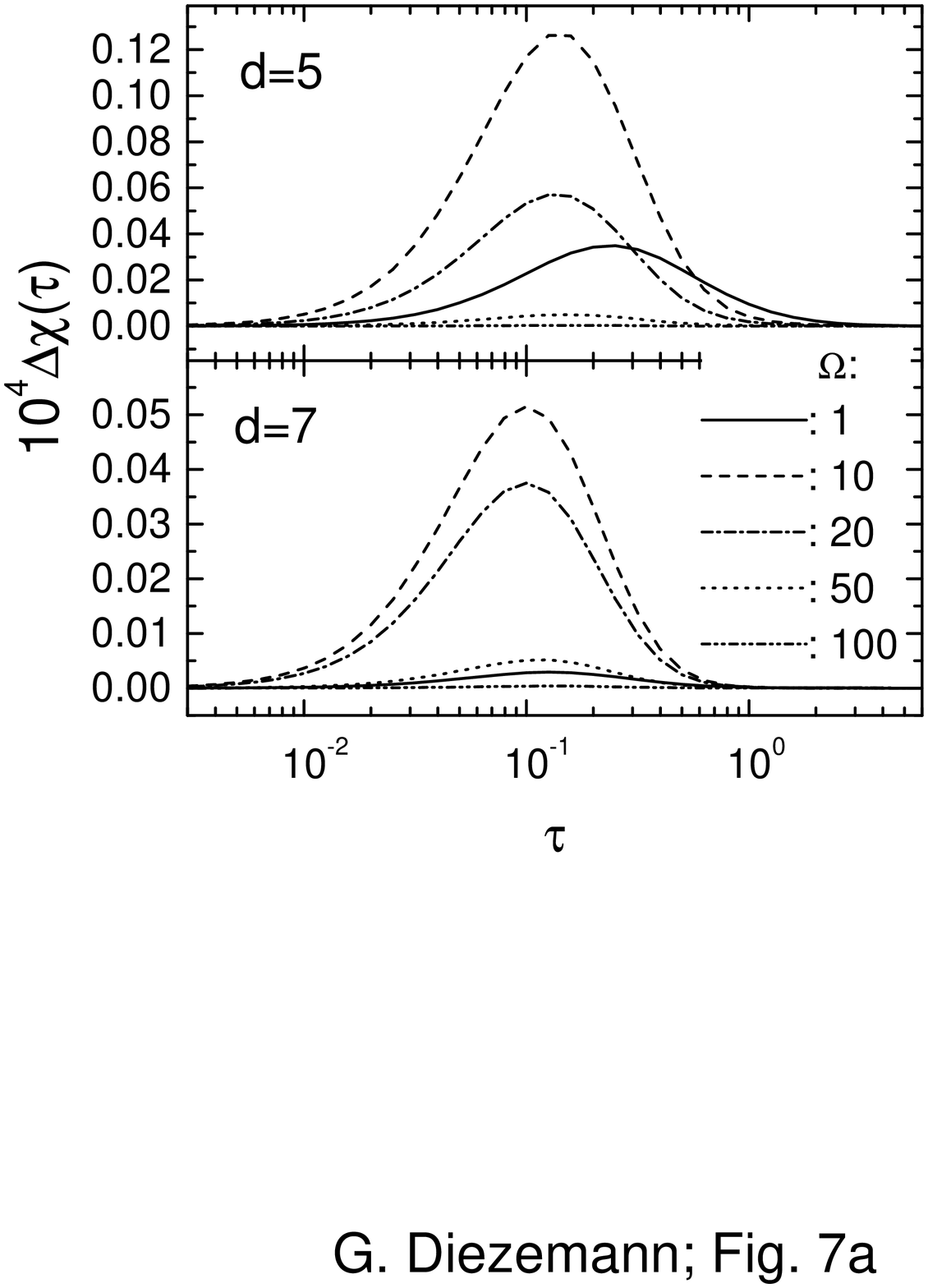}
\end{figure}
\newpage
\begin{figure}
\includegraphics[width=15cm]{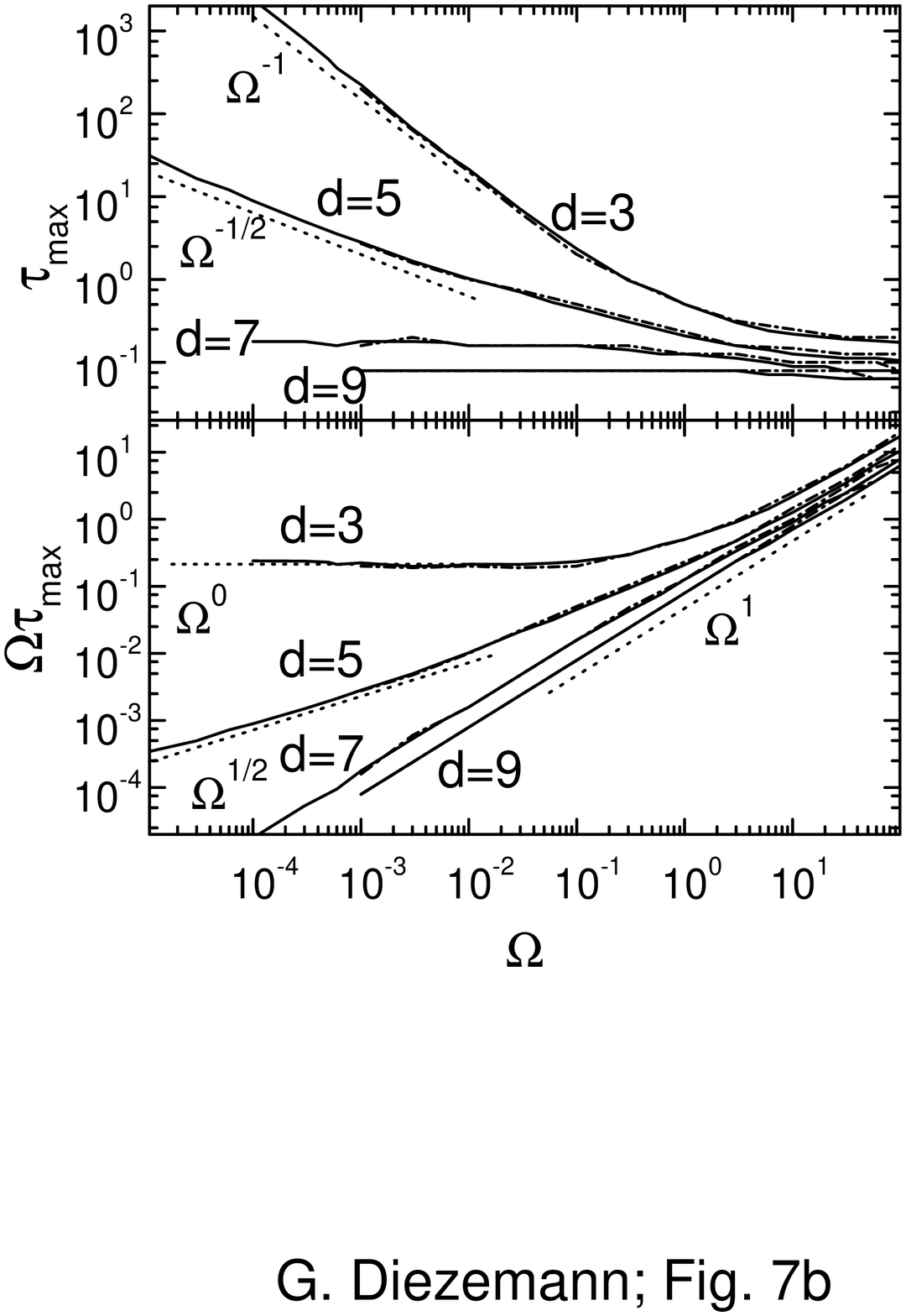}
\end{figure}
\newpage
\begin{figure}
\includegraphics[width=15cm]{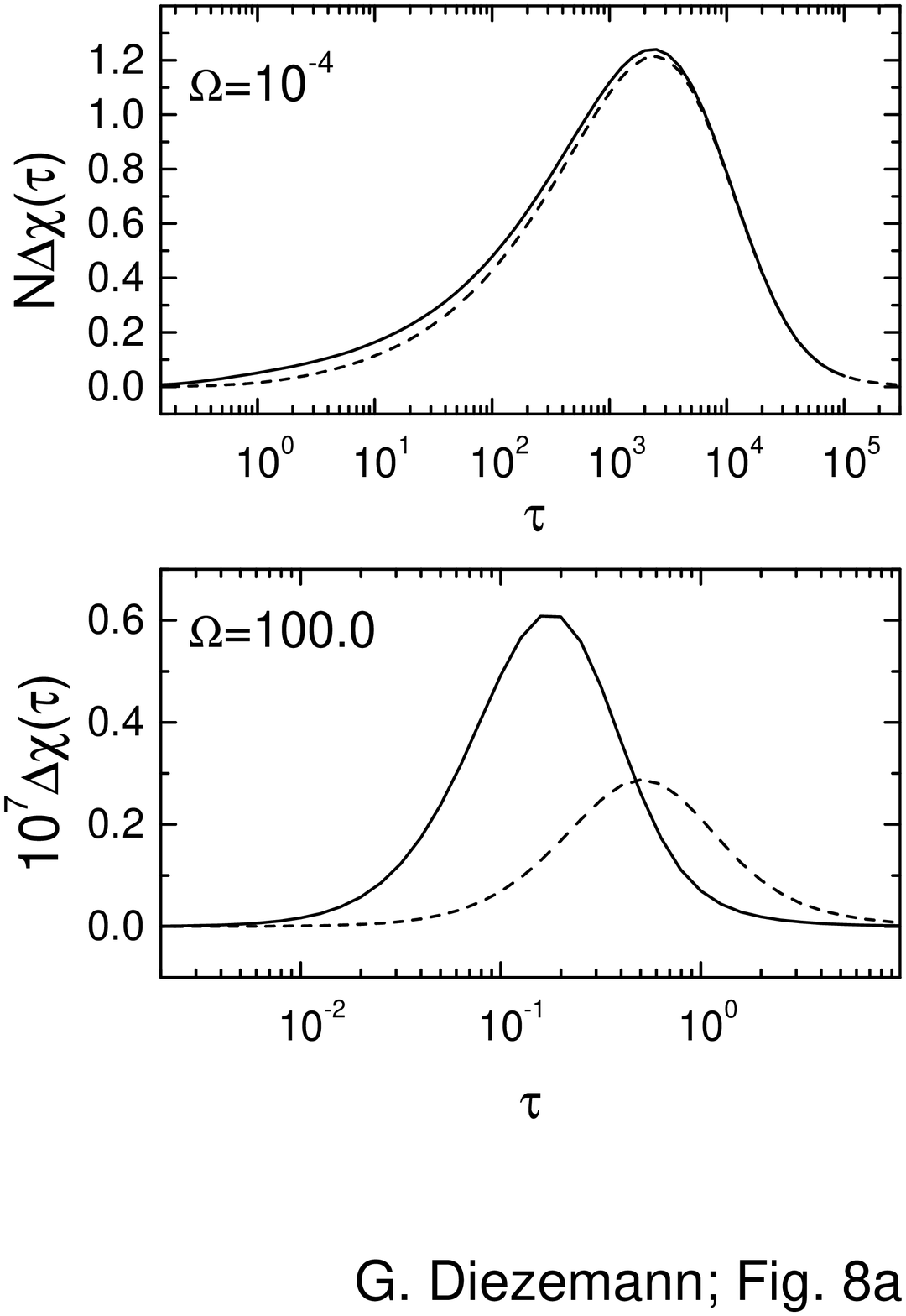}
\end{figure}
\newpage
\begin{figure}
\includegraphics[width=15cm]{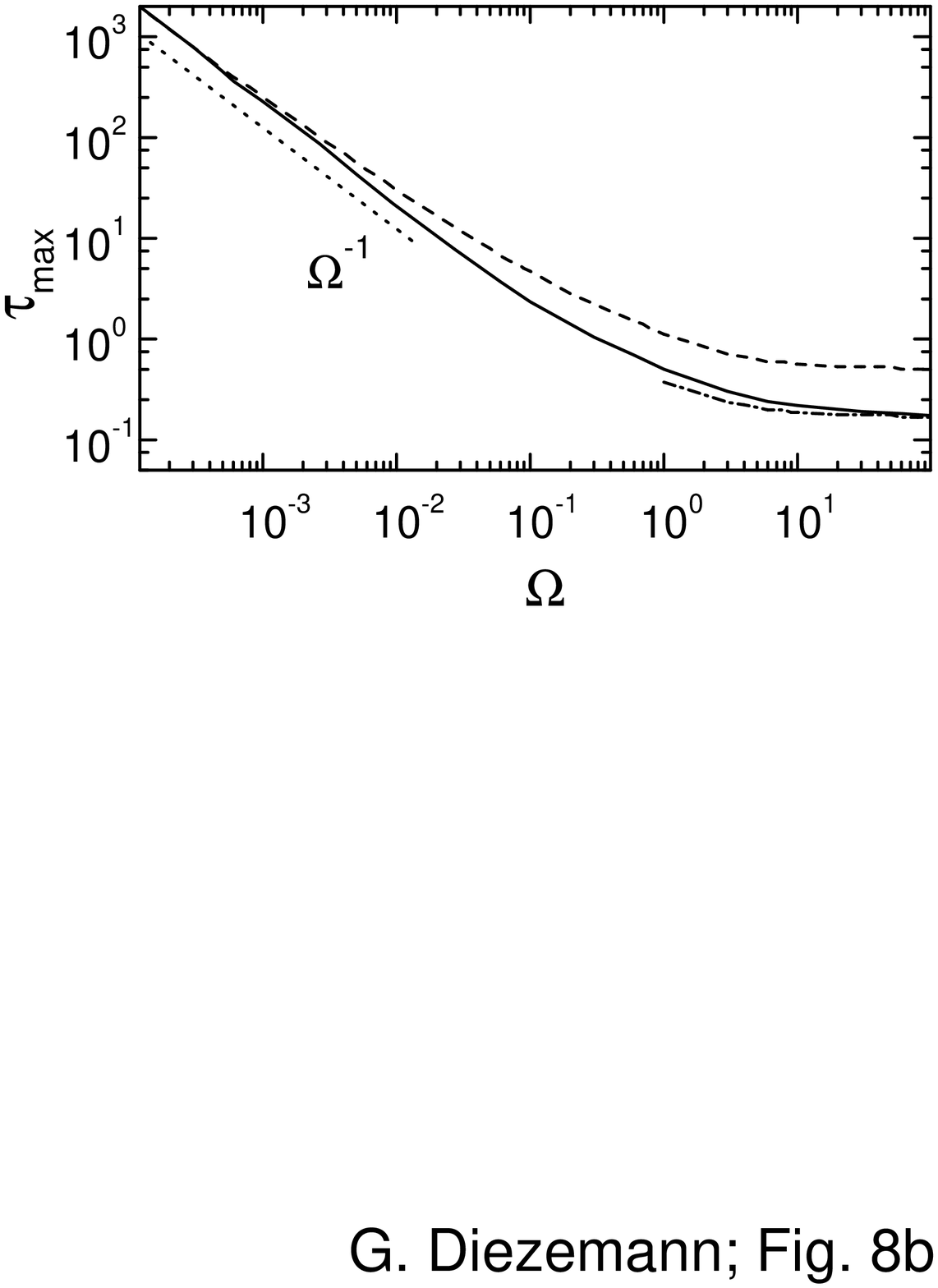}
\end{figure}
\end{document}